\documentclass[letterpaper,titlepage,11pt]{article}
\usepackage{hyperref}
\usepackage{amssymb,amsmath,amsfonts}
\usepackage{epsfig}
\def\clock{{\count0=\time
           \divide\count0 60
           \ifnum\count0<10 0\fi\the\count0
           \multiply\count0 -60 \advance\count0 \time
           :\ifnum\count0<10 0\fi \the\count0
         }}
\newcommand{\timestamp}{{\small\vbox{\hbox{\tt\jobname.tex}
\hbox{\the\day/\the\month/\the\year, \clock}}}}

\newcommand{\CF}{\mathcal{F}}

\newcommand{\CN}{\mathcal{N}}
\newcommand{\CO}{\mathcal{O}}

\newcommand{\CS}{\mathcal{S}}

\newcommand{\nn}{\nonumber}

\newcommand{\spa}{\ , \ \ }

\newcommand{\ds}{\displaystyle}

\newcommand{\ads}{\mbox{AdS}}

\numberwithin{equation}{section}

\begin{document}

\begin{titlepage}

\rightline{\vbox{\small\hbox{\tt NORDITA-2010-99} }}
 \vskip 1.8 cm

\centerline{\Huge \bf Heating up the BIon}
\vskip 1.5cm

\centerline{\large {\bf Gianluca Grignani$\,^{1}$},  {\bf Troels Harmark$\,^{2}$},}
\vskip 0.2cm \centerline{\large  {\bf Andrea Marini$\,^{1}$} ,  {\bf Niels A. Obers$\,^{3}$} and
{\bf Marta Orselli$\,^{3}$} }

\vskip 1.0cm

\begin{center}
\sl $^1$ Dipartimento di Fisica, Universit\`a di Perugia,\\
I.N.F.N. Sezione di Perugia,\\
Via Pascoli, I-06123 Perugia, Italy
\vskip 0.4cm
\sl $^2$ NORDITA\\
Roslagstullsbacken 23,
SE-106 91 Stockholm,
Sweden \vskip 0.4cm
\sl $^3$ The Niels Bohr Institute  \\
\sl  Blegdamsvej 17, DK-2100 Copenhagen \O , Denmark
\end{center}
\vskip 0.6cm

\centerline{\small\tt grignani@pg.infn.it, harmark@nordita.org, }
\centerline{\small\tt andrea.marini@fisica.unipg.it, obers@nbi.dk, orselli@nbi.dk}

\vskip 1.3cm \centerline{\bf Abstract} \vskip 0.2cm \noindent
We propose a new method to consider D-brane probes in thermal backgrounds. The method builds on the recently developed blackfold approach to higher-dimensional black holes. 
While D-brane probes in zero-temperature backgrounds are well-described by the Dirac-Born-Infeld (DBI) action, this method addresses how to probe thermal backgrounds. A particularly important feature is that the probe is in thermal equilibrium with the background. We apply our new method to study the thermal generalization of the BIon solution of the DBI action. The BIon solution is a configuration in flat space of a D-brane and a parallel anti-D-brane connected by a wormhole with F-string charge.
In our thermal generalization, we put this configuration in hot flat space.
We find that the finite temperature system behaves qualitatively different than its zero-temperature counterpart. In particular, for a given separation between the D-brane and anti-D-brane, while at zero temperature there are two phases, at finite temperature there are either one or three phases available.

\end{titlepage}

\small
\tableofcontents
\normalsize
\setcounter{page}{1}

\section{Introduction}

The Dirac-Born-Infeld (DBI) action is a low energy effective action for D-brane dynamics obtained by integrating out the massive open string degrees of freedom \cite{Fradkin:1985qd}. The first example in string theory where the full non-linear dynamics of the DBI action was exploited is that of the BIon solution \cite{Callan:1997kz,Gibbons:1997xz}. In the BIon solution the configuration in which an F-string ends on a point of a D-brane is resolved into a smooth solution of the DBI action where the F-string dissolves into the D-brane. 
The D-brane carries a world-volume electric flux which is the F-string charge. 
The new phenomena at play are that one can describe multiple coincident F-strings in terms of D-branes and furthermore that the F-strings go from being a one-dimensional object of zero thickness to be ``blown up" to a higher-dimensional brane wrapped on a sphere. These phenomena are captured thanks to the non-linear nature of the DBI action. Based on these phenomena, many important applications of the DBI action were found in the context of the AdS/CFT correspondence \cite{Maldacena:1997re}. For gravitons satisfying a BPS bound by moving on the equator of the $S^5$ of the $\ads_5\times S^5$ background it was found that they blow up to become Giant Gravitons, D3-branes wrapped on three-spheres, for sufficiently large energies \cite{McGreevy:2000cw,Grisaru:2000zn,Hashimoto:2000zp}.
Another interesting application is the Wilson loop, originally considered in the AdS/CFT correspondence using the Nambu-Goto F-string action \cite{Rey:1998ik,Maldacena:1998im}. Here, the ``blown up" version for a Wilson loop in a high-dimensional representation has been considered using the DBI action, either for the symmetric representation using a D3-brane \cite{Drukker:2005kx} or the antisymmetric representation using a D5-brane \cite{Yamaguchi:2006tq}.

The success of using the DBI action to describe D-branes probing zero-temperature backgrounds of string theory motivated the application of the DBI action as a probe of thermal backgrounds, particularly in the context of the AdS/CFT correspondence with either thermal AdS space or a black hole in AdS as the background \cite{Witten:1998zw}. Applications include meson spectroscopy at finite temperature, the melting phase transition of mesons and other types of phase transitions in gauge theories with fundamental matter (see \cite{Babington:2003vm} for early works on this). Furthermore, the thermal generalizations of the Wilson loop, the Wilson-Polyakov loop, in high-dimensional representations were considered \cite{Hartnoll:2006hr,Grignani:2009ua}.%
\footnote{Unlike in the D5-brane case, it seems that for the D3-brane case, which corresponds to a totally symmetric representation of the Polyakov loop, there is no solution \cite{Hartnoll:2006hr,Grignani:2009ua}.}

However, the validity of using the DBI action as a probe of thermal backgrounds is not clear. In general the equations of motion (EOMs) for any probe brane can be written as \cite{Carter:2000wv,Emparan:2009at}
\begin{equation}
\label{braneeoms}
K_{ab}{}^\rho T^{ab} = J \cdot F^\rho
\end{equation}
where $T_{ab}$ is the world-volume energy-momentum (EM) tensor for the brane, $K_{ab}{}^\rho$ is the extrinsic curvature given by the embedding geometry of the brane and the right hand side, $J \cdot F^\rho$, represents possible external forces arising from having a charged brane that couples to an external field. In the applications of the DBI action as a probe of thermal backgrounds the D-brane is treated as if the temperature of the background does not affect the physics on the brane. Therefore, the EM tensor that enters in the EOMs \eqref{braneeoms} is the same as in the zero-temperature case. However, there are degrees of freedom (DOFs) living on the brane that are ``warmed up" by the temperature of the thermal background, just like if one puts a cold finger in a big bathtub with warm water. The thermal background should thus act as a heat bath for the D-brane probe and the system should attain thermal equilibrium with the D-brane probe gaining the same temperature as the background. Because of the DOFs living on the brane this will change the EM tensor of the brane and thus in turn change the EOMs \eqref{braneeoms} that one should solve for the probe brane.

In this paper we study the thermal generalization of the BIon solution. This serves as a test case to study D-branes as probes of thermal backgrounds. The BIon solution is a solution of the DBI action for a D-brane probing ten-dimensional flat space-time, the D-brane world-volume having an electric flux interpreted in the bulk as an F-string. We shall instead consider ten-dimensional hot flat space as our background. The challenge is that one does not know what replaces the DBI action, which is a low energy effective action for a single extremal D-brane at weak string coupling, when turning on the temperature. However, in the regime of a large number $N$ of coinciding D-branes we have an effective description of the D-branes in terms of a supergravity solution in the bulk when $g_s N \gg 1$.
Using this supergravity description one can determine the EM tensor for the D-brane in the regime of large $N$. This EM tensor will then enable one to write down the EOMs \eqref{braneeoms} for a non-extremal D-brane probe in the regime of large $N$. This results in a new method in which one can replace the DBI action, which is a good description of a single D-brane probing a zero-temperature background, by another description that can describe $N$ coincident non-extremal D-branes probing a thermal background such that the probe is in thermal equilibrium with the background.

The idea behind the new method for non-extremal D-branes probing thermal backgrounds comes from the so-called blackfold approach
\cite{Emparan:2009at,Emparan:2007wm,upcoming} which recently has been developed in the study of higher-dimensional black holes. Indeed, the conceptual and technical developments of the blackfold approach to black holes provide the building blocks that we need in this paper. The blackfold approach provides a general description of black holes in a regime in which the black hole approximately is like a {\sl black} brane curved along a submani{\sl fold} of a background space-time (hence the name ``blackfold"). This regime entails in particular that the thickness of the black brane is much smaller than the length scale of the embedding geometry.%
\footnote{The brane thickness scale is given by the length scale over which the brane backreacts on the surrounding space-time.}
The general idea of the blackfold approach is thus that one integrates out the brane thickness scale and obtains an effective description in terms of the large scales over which the brane varies, such as the scale of the embedding geometry. Doing this one obtains an effective description with the world-volume EM tensor of the brane being that of a fluid and with the dynamical principle being the conservation of the EM tensor. The EOMs resulting from this consist of the conservation of the EM tensor on the world-volume along with the extrinsic EOMs for the blackfold of the form \eqref{braneeoms}.
 To leading order the brane can be regarded as a probe brane that does not backreact on the background geometry.%
\footnote{Obviously, if the distance to the brane is of order the brane thickness scale one would observe a backreaction to the surrounding geometry from the brane. However, we are precisely integrating out the brane thickness scale, thus the probe approximation assumes in this case that the backreaction from the brane is sufficiently small to be neglible for the large scale physics described by the extrinsic EOMs \eqref{braneeoms}. $I.e.$ only when considering corrections to the probe approximation one should start taking into account the backreaction of the brane on the background geometry when computing the extrinsic curvature tensor $K_{ab}{}^\rho$.}
Thus, as we shall see, one can parallel the probe approximation in the blackfold approach with the probe approximation that the DBI action assumes and the only difference in the extrinsic EOMs \eqref{braneeoms} is that one should replace the DBI EM tensor with that of the fluid EM tensor for the black brane.

Before turning to finding the thermal generalization of the BIon solution, we begin this paper in Section \ref{sec:bionsolution} by reviewing the original BIon solution. As part of this we introduce the physical set up of the BIon such as how we embed the BIon solution in flat ten-dimensional space-time. We restrict ourselves in this paper to consider D3-branes. We review that the BIon solution either takes the form of an infinite spike with an F-string charge coming out of the D3-brane, describing a number of coincident F-strings ending on the D3-brane, or the form of a D3-brane and a parallel anti-D3-brane connected by a ``wormhole" with F-string charge. The latter solution would correspond in the linearized regime to a string stretched between a D3-brane and an anti-D3-brane.

In Section~\ref{sec:heatingDBI}  we consider the question of D-brane probes in thermal backgrounds in general. We approach this in Section \ref{sec:extrinsic} by first reconsidering the general form of the EOMs for the DBI action and show that they can be cast in the form \eqref{braneeoms} with the world-volume EM tensor being given from the DBI action. Since \eqref{braneeoms} are the general EOMs for a probe brane all we need to do in order to obtain the EOMs for a thermal D-brane probe is to find the corresponding EM tensor. This we attack in Section \ref{sec:EMblackbrane}. Since the BIon solution includes an F-string charge on the brane we find the EM tensor for the non-extremal D3-F1 brane bound state. We subsequently explain how this can be used to write down the EOMs for a thermal D-brane probe by employing the blackfold approach. We find in particular that the EOMs for the thermal D-brane probe in the extremal limit are the same as for the DBI action probe, if one sets the number of D3-branes $N$ to 1, while keeping the ratio of the F-string charge relative to the D3-brane charge fixed. Finally, in Section \ref{sec:argument} we give a detailed argument why a D-brane probe in a thermal background is not accurately described by the DBI action.

In Section~\ref{sec:solution} we consider the BIon solution in the background of hot flat space. We first write down the EOMs \eqref{braneeoms} explicitly for the setup for the BIon solution introduced in Section \ref{sec:bionsolution}. This results in a single equation for the profile of the solution.
Based on the blackfold approach, we propose an action from which the EOMs can be derived. For our setup this gives the same equation as before, but written in a form that makes it readily solvable. Using this we find the general solution for the D3-F1 blackfold configuration that generalizes the BIon solution of the DBI action to non-zero temperature.

In Section~\ref{sec:braneseparation} we examine more closely the brane-antibrane wormhole solution at finite temperature. This solution consists of $N$ D3-branes and $N$ anti-D3-branes, separated by a distance $\Delta$, connected by an F-string charged wormhole with minimal sphere radius $\sigma_0$. We consider $\Delta$ as a function of $\sigma_0$  when $\sigma_0$ is varied for a given temperature and compare in this way our configuration at non-zero temperature with the corresponding extremal configuration. At zero temperature, the wormhole solution is characterized by a thin branch with small $\sigma_0$ and a thick branch with large $\sigma_0$ for fixed $\Delta$. We discover that, when the temperature is turned on, the separation distance $\Delta$ between the brane-antibrane system develops a local maximum in the region corresponding to the zero temperature thin branch. This is a new feature compared to the zero temperature case. The existence of this maximum gives rise to three possible phases with different $\sigma_0$ for a given $\Delta$. For small temperatures and/or large $\sigma_0$ the $\Delta$ as a function of $\sigma_0$ resembles increasingly closely the zero temperature counterpart.

We end in Section \ref{sec:concl} with the conclusions and discussions of further directions. 

In a forthcoming paper \cite{forthcoming} we investigate the thermodynamics of our newly found solution by analyzing the physics of the three different phases found in Section \ref{sec:braneseparation} in the brane-antibrane wormhole configuration and by examining whether it is possible to construct a thermal generalization of the infinite spike solution.

\section{BIon solution of DBI revisited}
\label{sec:bionsolution}

At low energies, the effective theory describing D-brane dynamics is given by the DBI action.
The DBI theory is a non-linear theory of a $U(1)$ gauge field living on the world-volume and a set of neutral scalars describing transverse fluctuations.
In the linearized regime, this reduces in particular to Maxwell electrodynamics for
 the gauge field while the scalars are free and massless.
In this linear approximation there exist world-volume solutions describing
Maxwell point charges with delta function sources. In the bulk theory
these are interpreted as F-strings ending on a point charge on the D-brane.
However, this picture is changed when one takes into account the
non-linearities of the DBI action. It has been found that
the brane curves before one reaches the point charge. The solution of DBI theory describing this is
called a BIon, and its bulk interpretation is that of F-strings dissolving into the D-brane \cite{Callan:1997kz,Gibbons:1997xz}.
Furthermore, these BIons solutions can describe a
configuration of two parallel D-branes (with one being an anti-brane)
connected by F-strings.

In this section we review the BIon solution since our aim in this paper is to describe its thermal generalization.
This will also enable us to introduce
the setup and provide the results to which we can compare the thermal
solution in the zero temperature limit. We will first briefly
review how the solution is obtained from the DBI action using the Hamiltonian
approach and subsequently describe the corresponding spike and wormhole solutions.

We consider in this paper only the D3-brane BIon solution, and its thermal generalization. However, all our results and considerations can readily be extended to general D$p$-branes.

\subsection{DBI action and setup \label{sec:DBI}}

Consider a D3-brane embedded in a 10-dimensional space-time with the only background flux turned on being the Ramond-Ramond five-form flux.
 We furthermore assume the background to have a constant dilaton.
The DBI action for the D3-brane then takes the form
\begin{equation}
\label{DBIaction}
I_{\rm DBI} = - T_{\rm D3} \int_{\rm w.v.} d^4 \sigma \sqrt{ - \det( \gamma_{ab} + 2\pi l_s^2 F_{ab} )} + T_{\rm D3} \int_{\rm w.v.} P [ C_{(4)} ]
\end{equation}
where the integrals are performed over the four-dimensional world-volume. Here we have defined the induced world-volume metric
\begin{equation}
\label{gamma}
\gamma_{ab} = g_{\mu\nu} \partial_a X^\mu \partial_b X^\nu
\end{equation}
where $g_{\mu\nu}$ is the background metric, $X^\mu(\sigma^a)$ is the embedding of the brane in the background with $\sigma^a$ being the world-volume coordinates, $a,b=0,1,2,3$ are world-volume indices and $\mu,\nu=0,1,...,9$ are target space indices. Furthermore, $F_{ab}$ is the two-form field strength living on the D-brane, $C_{(4)}$ is the RR-four form gauge field of the background and $P [ C_{(4)} ]$ is its pull-back to the world-volume. Finally, the D3-brane tension is $T_{\rm D3} = [ (2\pi)^3 g_s l_s^4 ]^{-1}$ where $g_s$ is the string coupling and $l_s$ is the string length.

\subsubsection*{Embedding}

To describe the BIon we specialize to an embedding of
the D3-brane world volume in 10D Minkowski space-time with metric
\begin{equation}
\label{background}
   ds^2=-dt^2+ dr^2 +r^2 ( d\theta^2+ \sin ^2\theta d\phi ^2) +\sum_{i=1}^6 dx_i^2 \, .
\end{equation}
without background fluxes.
Choosing the world volume coordinates of the D3-brane as $\{\sigma^a,\ a=0 \dots 3\}$ and
defining $\tau \equiv \sigma^0$, $\sigma \equiv \sigma^1$, the embedding of the three-brane is given
by
\begin{equation}
\label{embedding}
t (\sigma^a )= \tau  \spa r(\sigma^a ) = \sigma  \spa  x_1 (\sigma^a )= z(\sigma)
\spa
\theta (\sigma^a ) = \sigma^2 \spa \phi(\sigma^a ) = \sigma^3 \ .
\end{equation}
and the remaining coordinates $x_{i=2..6}$ are constant. There is thus one non-trivial embedding
function $z(\sigma)$ that describes the bending of the brane.
The induced metric on the brane is then
\begin{equation}\label{flatindmetric}
\gamma_{ab} d\sigma^a d\sigma^b =  -
d\tau^2 + \left( 1+ {z'(\sigma)}^2 \right)  d\sigma^2 +
\sigma^2 \left( d\theta^2 + \sin^2 \theta d\phi^2 \right)\, .
\end{equation}
so that the spatial volume element is
$dV_{(3)}= \sqrt{ 1 + z'(\sigma)^2} \sigma^2 d \Omega_{(2)}$.

To get the appropriate F-string flux on the brane we turn on the world-volume gauge field strength component $F_{01}$. With this, the DBI action \eqref{DBIaction} gives the following Lagrangian
\begin{equation}
\label{DBIaction2}
L = - 4 \pi T_{\rm D3}  \int_{\sigma_0}^\infty d\sigma \sigma^2 \sqrt{ 1 + {z'(\sigma)}^2 - ( 2\pi l_s^2 F_{01} )^2 }
\end{equation}
Note that we assumed $F_{01}$ to depend only on $\sigma$ since this is required for spherical symmetry.

\subsubsection*{Boundary conditions}

\begin{figure}[h!]
\centerline{\includegraphics[scale=0.5]{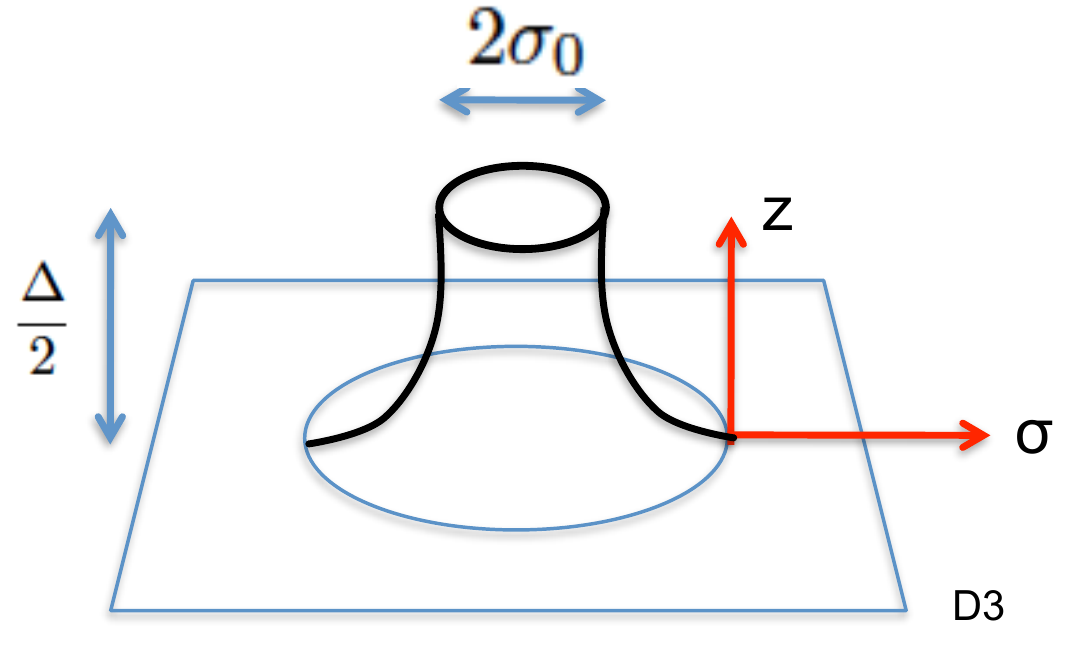}}
\caption{\small Illustration of the setup, showing the embedding function $z(\sigma)$ and
the definition of the parameters $\sigma_0$ and $\Delta$.}
\label{fig:setup}
\end{figure}

We have two boundary conditions on the BIon solution. The first one is
\begin{equation}
\label{sigmainfinity}
z(\sigma)\rightarrow 0 \ \ \mbox{for} \ \ \sigma \rightarrow \infty \, .
\end{equation}
This condition ensures that far away from the center at $r=0$ the D3-brane is flat and infinitely extended with $x_1=0$. Decreasing $\sigma$ from $\infty$ the brane has a non-trivial profile $x_1=z(\sigma)$. In general we have a minimal sphere with radius $\sigma_0$ in the configuration. For a BIon geometry $z(\sigma)$ is naturally a decreasing function of $\sigma$ where at $\sigma=\sigma_0$ the function $z(\sigma)$ reaches its maximum. Thus, $\sigma$ takes values in the range from $\sigma_0$ to $\infty$. At $\sigma_0$ we impose a Neumann boundary condition
\begin{equation}
\label{sigma0deriv}
z'(\sigma) \rightarrow -\infty \ \ \mbox{for} \ \ \sigma \rightarrow \sigma_0 \, .
\end{equation}
The rationale of this condition is that if $z(\sigma_0) < \infty$ the brane system cannot end at $(r=\sigma_0,x_1=z(\sigma_0))$ because of charge conservation and this boundary condition, as we describe below, enables us to attach a mirror of the solution, reflected in the hyperplane $x_1 = z(\sigma_0)$. In line with this, we define
\begin{equation}
\label{Delta}
\Delta \equiv 2 z(\sigma_0)
\end{equation}
as the separation distance between the brane and its mirror. Fig.~\ref{fig:setup} illustrates our setup and the definitions of $\sigma_0$
and $\Delta$.

\subsection{BIon solution}
\label{sec:extremalwormhole}

\begin{figure}[h!]
\centerline{\includegraphics[scale=1]{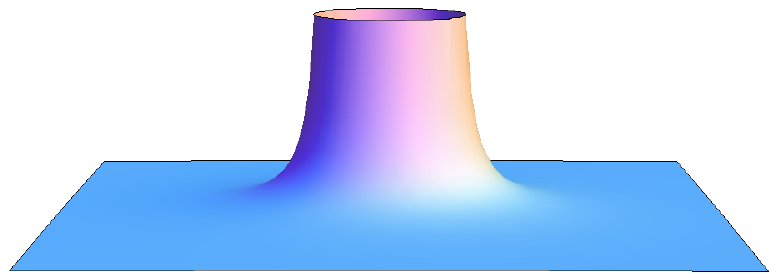}}
\caption{\small Sketch of solution with finite size throat.}
\label{fig:wh_bottom}
\end{figure}

 We now consider the Hamiltonian corresponding to the  Lagrangian \eqref{DBIaction2}.
 To derive this we need the canonical momentum density $4\pi \sigma^2 \Pi(\sigma) =
 \delta {L}/\delta (\partial_\tau A_1)$
associated with the world-volume gauge field component $A_1$.
 Using ${F_{01}}=\partial_\tau {A_1}-\partial_\sigma A_0$ this gives
\begin{equation}\label{Pi}
 \Pi(\sigma)=T_{\rm D3}\,\frac{(2\pi l_s^2)^2 F_{01}}
   {\sqrt{1+ {z'}^2 - ( 2\pi l_s^2 F_{01} )^2}}
\end{equation}
so that the Hamiltonian can be easily constructed as
\begin{equation}\label{CMconstr}
  H_{\rm DBI} = 4\pi \int d\sigma \sigma^2 \Pi(\sigma)\partial_\tau {A_1(\sigma,\tau)}-L=4\pi\int d\sigma
  \left[  \sigma^2
  \Pi(\sigma)F_{01} - \partial_\sigma (\sigma^2 \Pi(\sigma)) A_0(\sigma)\right]-L
\end{equation}
Here we have in the second step integrated by parts the term proportional to $\partial_\sigma A_0$, showing that  $A_0$ can be considered as a Lagrange multiplier imposing the constraint $\partial_\sigma
(\sigma^2\Pi(\sigma))=0$ on the canonical momentum.  Solving this constraint
gives
\begin{equation}\label{gauss}
\Pi(\sigma)=\frac{k}{4\pi \sigma^2}=\frac{T_{\rm D3}\kappa}{\sigma^2 T_{\rm F1}}
\end{equation}
where $k$ is an integer, $T_{\rm F1}=1/2\pi l_s^2$ is the tension of a fundamental string
and  we have defined $\kappa  \equiv \frac{k T_{\rm F1}}{4\pi  T_{\rm D3}}
= k \pi g_s l_s^2 $.
Using \eqref{gauss} in \eqref{CMconstr} the Hamiltonian becomes
\begin{equation}\label{CMham}
H_{\rm DBI} =  4\pi  T_{\rm D3}\int d\sigma
  \sqrt{1+z'(\sigma)^2}  F_{\rm DBI} (\sigma) \spa F_{\rm DBI} (\sigma) \equiv \sigma^2 \sqrt{1 + \frac{\kappa^2}{\sigma^4}}
\end{equation}
The resulting EOM for ${z(\sigma)}$, obtained by varying \eqref{CMham}, is
\begin{equation}
\label{theeom0}
\left( \frac{z'(\sigma) F_{\rm DBI} (\sigma)}{\sqrt{1+z'(\sigma)^2 }} \right)' = 0
\end{equation}
Solving for $z(\sigma)$ subject to the boundary conditions stated above,
we obtain
\begin{equation}\label{Xprime}
   -z'(\sigma)= \left( \frac{F_{\rm DBI} (\sigma)^2}{F_{\rm DBI} (\sigma_0)^2} -1 \right)^{-\frac{1}{2}}  = \frac{\sqrt{\sigma_0^4+\kappa^2}}{\sqrt{\sigma^4-\sigma_0^4}}
\end{equation}
where we recall that $\sigma_0$ is the minimum value of the two-sphere
radius $\sigma$. The explicit solution for $z$ can be obtained by integrating the expression in \eqref{Xprime}
\begin{equation}\label{X}
   z(\sigma)=\int_\sigma^{\infty}d\sigma'\frac{\sqrt{\sigma_0^4+\kappa^2}}{\sqrt{\sigma'{}^4-\sigma_0^4}}
\end{equation}
In particular for non-zero
$\sigma_0$ this represents a solution with a finite size throat, as illustrated in Fig.~\ref{fig:wh_bottom}.

From the expression in \eqref{Xprime} we can compute the energy density by evaluating the integrand
of \eqref{CMham}, yielding
\begin{equation}\label{Hden1}
   \frac{dH}{d\sigma}=4\pi\,T_{\rm D3}
  \sqrt{\left(1+z'(\sigma)^2\right)\left(\sigma^4+\kappa^2\right)}=4\pi\,T_{\rm D3}\frac{\sigma^4+\kappa^2}{\sqrt{\sigma^4-\sigma_0^4}}
\end{equation}
From this, dividing by the derivative of the solution $z'(\sigma)$, we compute
the energy density along the brane
\begin{equation}
\label{energydensity}
 \frac{dH}{dz} = \frac{1}{z'(\sigma)}\frac{dH}{d\sigma}=4\pi\,T_{\rm D3}\frac{\sigma^4+\kappa^2}{\sqrt{\sigma_0^4+\kappa^2}} .
\end{equation}
which is finite for $\sigma$ in the range $[\sigma_0,\infty)$.

\subsubsection*{Spike solution}

\begin{figure}[h!]
\centerline{\includegraphics[scale=1]{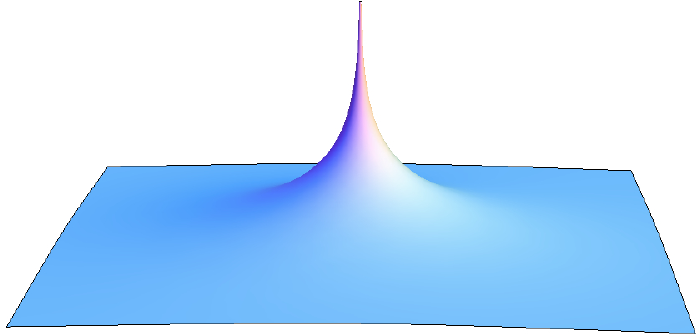}}
\caption{\small Sketch of the spike configuration.}
\label{fig:spikewormhole}
\end{figure}

For $\sigma_0=0$ the integral in Eq.~\eqref{X} gives the Coulomb-charge type of solution
\begin{equation}
z(\sigma)=\frac{\kappa}{\sigma}
\end{equation}
{\it i.e.} the spike solution (see Fig.~\ref{fig:spikewormhole}).
In Ref.~\cite{Callan:1997kz} it was shown that the energy corresponding to this solution is the energy of a fundamental string of a given length. This was done by comparing the integral appearing in \eqref{CMham}  with the explicit form of the solution $z(\sigma)$ at a point near the end of the spike, in the linear approximation and with a suitable regularization of the integral providing the energy.
In the non-linear case, and also to avoid divergences, it is more
convenient to compute the energy density along the brane as in  \eqref{energydensity}.
In particular setting $\sigma=\sigma_0=0$ in \eqref{energydensity} we find that
the energy density at the tip of the spike is given by
\begin{equation}
   \left.\frac{dH}{dz}\right|_{\sigma=\sigma_0=0}=4\pi\,T_{\rm D3}\kappa=k T_{\rm F1}
\end{equation}
where we used $\kappa$ defined below \eqref{gauss}.
We thus find that this is the tension of the fundamental string
times the number of strings $k$, as expected. 

\subsubsection*{Wormhole solution}

\begin{figure}[h!]
\centerline{\includegraphics[scale=1]{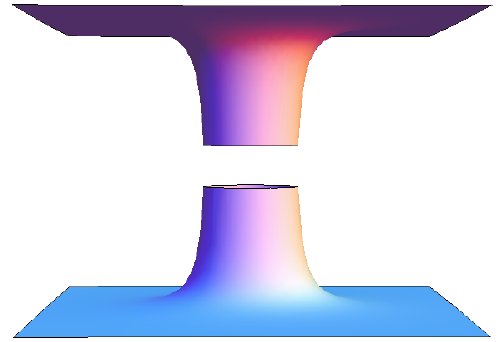}\hskip 1cm \includegraphics[scale=1]{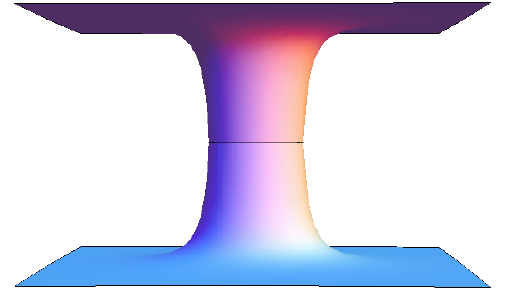}}
\caption{\small Attaching a mirror solution to construct a wormhole configuration.}
\label{fig:wh}
\end{figure}

For $\sigma_0=0$ we showed above that the solution \eqref{X} corresponds to a spike. However, as explained in Sec.~\ref{sec:DBI} for more general values of $\sigma_0$ one can use the solution
to construct  a configuration  representing strings going between branes and anti-branes~\cite{Callan:1997kz}, to which we refer as the wormhole configuration (see Fig.~\ref{fig:wh}).

Obviously, a system of a D-brane separated from an anti-D-brane is unstable since the branes attract each other both gravitationally and electrically. However, the time scale of this is very large for small string coupling $g_s \ll 1$ since the tension of a D-brane goes like $1/g_s$ while the gravitational coupling goes like $g_s^2$. Indeed, for a Dp-brane and anti-Dp-brane system the time scale $t$ is of order
\begin{equation}
\frac{t}{l_s} \sim \frac{1}{\sqrt{g_s}} \left( \frac{\Delta}{l_s} \right)^{\frac{9-p}{2}}
\end{equation}
where $\Delta$ is the distance between the brane and anti-brane. 

\begin{figure}[ht]
\centering
\includegraphics[scale=1]{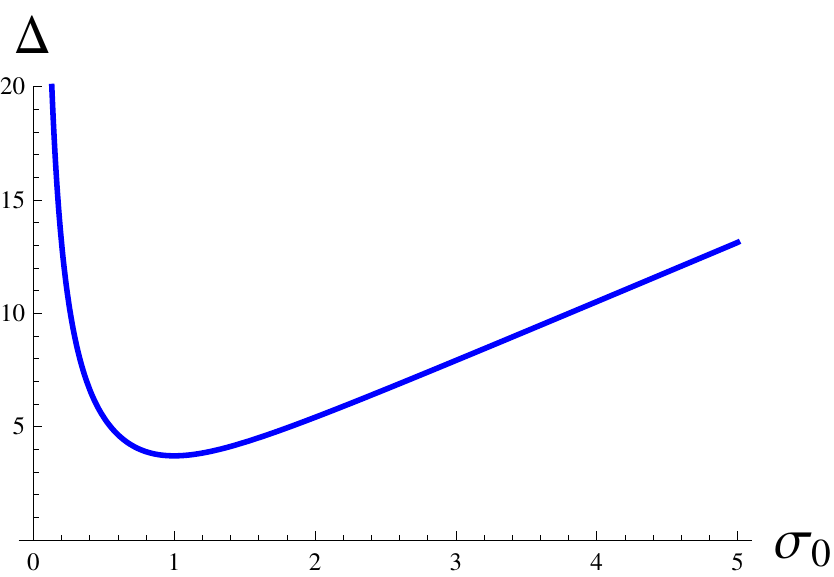}
\caption{{\small $\Delta$ for $\kappa=1$ in the Callan-Maldacena case}
\label{DeltaCM}
}
\end{figure}

Turning back to the wormhole configuration, the separation \eqref{Delta}  between the D3-brane and anti-D3-brane can be computed from \eqref{X} and is given by
\begin{equation}\label{separation}
   \Delta\equiv 2z(\sigma_0)=\frac{2\sqrt{\pi}\Gamma(\frac{5}{4})\sqrt{\sigma_0^4+\kappa^2}}{\Gamma(\frac{3}{4})\sigma_0}
\end{equation}
A plot of this quantity as a function of $\sigma_0$ is given in Fig.\ref{DeltaCM}. It is clear that there is a minimum value of the distance between the two branes, the minimum occurs at $\sigma_0=\sqrt{\kappa}$ and its value is
\begin{equation}\label{DeltaminCM}
\Delta_{\rm min}=\frac{2 \sqrt{2 \pi } \Gamma \left(\frac{5}{4}\right) \sqrt{\kappa }}{\Gamma
   \left(\frac{3}{4}\right)}
\end{equation}
Since $\kappa$ is related to the world-volume gauge field, we see that only for zero electric field the two branes can annihilate.
For large $\sigma_0$ the distance $\Delta$ between the two branes grows linearly with $\sigma_0$.
We can now solve \eqref{separation} for $\sigma_0$ by keeping fixed the distance between the branes $\Delta$ and the number of strings, which is done by keeping the charge parameter $\kappa$ fixed. We obtain
\begin{equation}\label{r02}
   \sigma_0^2=\frac{\Delta ^2\pm\sqrt{\Delta ^4-4 a^4 \kappa ^2}}{2 a^2}
   \end{equation}
where the numerical constant $a$ is given by $a^2=\frac{2\sqrt{\pi } \Gamma \left(\frac{5}{4}\right)}{\Gamma \left(\frac{3}{4}\right)}$.
There are  two solutions which, for large $\Delta$, behave as~\cite{Callan:1997kz}
\begin{equation}
\label{largeDelta}
\sigma_0\simeq \frac{a\kappa}{\Delta}~,~~~~~~~~\sigma_0\simeq\frac{\Delta}{a}
\end{equation}
In the first case, the ``thin throat" branch, the radius of the throat goes to zero as $\Delta\to\infty$. In the second case, the ``thick throat" branch, the radius of the throat grows linearly with $\Delta$ as $\Delta \rightarrow \infty$.

\section{Heating up DBI solutions by the blackfold approach}
\label{sec:heatingDBI}

Our aim in the rest of the paper is to turn on the temperature and analyze
what happens to the BIon solution described in the previous
section. We propose in this section that the appropriate
framework in which to heat up DBI solutions is provided by the blackfold approach
\cite{Emparan:2009at,Emparan:2007wm}. We will motivate this here by first casting
the DBI EOMs in a form that makes apparent the generalization
to the thermally excited case, as discussed in Section \ref{sec:extrinsic}. The latter involves the EM tensor for a black D3-F1 brane bound state, which is subsequently obtained in Section \ref{sec:EMblackbrane}.

In Section \ref{sec:argument} we give a detailed argument for why our thermal D-brane probe, based on the blackfold approach, should give a more accurate way to probe finite temperature backgrounds in string theory than the Euclidean DBI probe method that has been used in the literature.

\subsection{Extrinsic embedding equations from DBI action}
\label{sec:extrinsic}

In this section we write a general expression of the EOMs for the DBI action. This takes the form of a set of extrinsic embedding equations. Below we shall use this to connect to the blackfold approach of \cite{Emparan:2009at,Emparan:2007wm} and thence to go to the thermal case.

Our starting point%
 \footnote{Our considerations are easily generalized to D$p$-branes}
 is the DBI action for the D3-brane given in \eqref{DBIaction}.
Before considering the EOMs we first obtain the world-volume EM tensor. This can be done by varying the action \eqref{DBIaction} with respect to the world-volume metric $\gamma_{ab}$ in \eqref{gamma}, $i.e.$
\begin{equation}
T^{ab} = \frac{2}{\sqrt{\gamma}}     \frac{\delta I_{\rm DBI}}{\delta \gamma_{ab}}
\end{equation}
We find
\begin{equation}
\label{EMDBI}
T^{ab} =  - \frac{T_{\rm D3}}{2} \frac{\sqrt{-\det
(\gamma+  2\pi l_s^2 F)}}{\sqrt{\gamma }} \left[ ((\gamma+  2\pi l_s^2 F)^{-1})^{ab} +
((\gamma+ 2\pi l_s^2 F)^{-1})^{ba} \right]
\end{equation}
where we defined the determinant $\gamma = - \det (\gamma_{ab} )$.

We now consider the EOMs for the DBI action \eqref{DBIaction}. These are found by variation of the embedding map $X^\mu(\sigma^a)$. We first notice that the Born-Infeld term in \eqref{DBIaction} only contributes through the variation of the world-volume metric. We compute
\begin{equation}
\delta \gamma_{ab} =   g_{\mu\nu,\lambda}  \partial_a X^\mu \partial_b X^\nu \delta
X^\lambda + g_{\mu\lambda} ( \partial_a X^\mu \partial_b
\delta X^\lambda + \partial_b X^\mu \partial_a \delta
X^\lambda)
\end{equation}
Hence we can write the variation of the Langrangian density of \eqref{DBIaction} as
\begin{align}
&\frac{1}{2} \sqrt{\gamma} T^{ab}  (   g_{\mu\nu,\lambda}  \partial_a X^\mu \partial_b X^\nu \delta
X^\lambda +2 g_{\mu\lambda} \partial_a X^\mu \partial_b
\delta X^\lambda ) \nn \\ & + \frac{T_{\rm D3}}{4!} \epsilon^{abcd} \partial_a X^\mu \partial_b X^\nu \partial_c X^\rho ( 4 \partial_d \delta X^\lambda C_{\mu\nu\rho\lambda} + \partial_d X^\alpha C_{\mu\nu\rho \alpha, \lambda }  \delta X^\lambda )
\end{align}
giving the EOMs
\begin{eqnarray}
\label{preEOM} &&
\frac{1}{2}  \sqrt{\gamma} T^{ab}
g_{\mu\nu,\lambda} \partial_a X^\mu \partial_b
X^\nu -
\partial_b ( \sqrt{\gamma} T^{ab} ) g_{\mu\lambda}
\partial_a X^\mu  -  \sqrt{\gamma} T^{ab} g_{\mu\lambda,\nu} \partial_a X^\mu \partial_b X^\nu \nn \\ && - \sqrt{\gamma} T^{ab} g_{\mu\lambda} \partial_a \partial_b X^\mu  + \frac{T_{\rm D3}}{4!} \epsilon^{abcd} \partial_a X^\mu \partial_b X^\nu \partial_c X^\rho \partial_d X^\alpha (C_{\mu\nu\rho\alpha,\lambda} - 4 C_{\mu\nu\rho\lambda,\alpha}  ) = 0
\end{eqnarray}
We define now the projector $h^{\mu\nu}$ along the tangent directions to the D3-brane
\begin{equation}
h^{\mu\nu} = \gamma^{ab} \partial_a X^\mu \partial_b X^\nu
\end{equation}
along with the projector $\perp^{\mu\nu}$ along the orthogonal directions, defined as $\perp^{\mu\nu} = g^{\mu\nu} - h^{\mu\nu} $. Using these we can define the extrinsic curvature tensor for the embedding
\begin{equation}
\label{extrcurv}
K_{ab}{}^\rho = \perp^\rho {}_\lambda ( \partial_a \partial_b X^\lambda + \Gamma^\lambda_{\mu\nu} \partial_a X^\mu \partial_b X^\nu )
\end{equation}
See \cite{Emparan:2009at} for more on these geometrical quantities.
We define furthermore the partial pullback of the RR five-form field strength $F_{(5)} = dC_{(4)}$
\begin{equation}
F_{\lambda abcd} = \partial_a X^\mu \partial_b X^\nu \partial_c X^\rho \partial_d X^\alpha F_{\lambda \mu\nu\rho\alpha }
\end{equation}
and the D3-brane RR charge current
\begin{equation}
\label{RRcurrent}
J^{abcd} = T_{\rm D3} \frac{1}{\sqrt{\gamma}} \epsilon^{abcd}
\end{equation}
Projecting now the EOMs \eqref{preEOM} with $\perp^{\rho\lambda}$ we can write the resulting EOMs as
\begin{equation}
\label{extrinsic}
T^{ab} K_{ab}{}^\rho =  \perp^{\rho \lambda} \frac{1}{4!}
J^{abcd} F_{\lambda abcd}
\end{equation}
using the above definitions. This is the extrinsic equation for the D3-brane, with the EM tensor given by \eqref{EMDBI}. As explained in \cite{Emparan:2009at} the equation \eqref{extrinsic} is basically the Newton's second law of brane mechanics, with $T^{ab}$ replacing the mass and $K_{ab}{}^\rho$ the acceleration of the point particle, while the right hand side generalizes  the Lorentz force for a charged particle.%
\footnote{If we instead project the EOMs  \eqref{preEOM} with $h^{\rho\lambda}$ we obtain the equation for conservation of the EM tensor $T_{ab}$ on the brane. See \cite{Emparan:2009at}  for general comments on this.}

The significance of rewriting the EOM's of the DBI action as  \eqref{extrinsic} is that the generalization to a thermally excited D3-brane becomes more apparent. In fact, as we shall see below, in some sense all we need to do is to replace the EM tensor \eqref{EMDBI} with that of a thermally excited brane in \eqref{extrinsic} and then we have the EOMs for a thermally excited D3-brane. However, in order to do this we have to shift to a regime in which the EM tensor of a thermally excited D3-brane is known: When we have a large number $N$ of coincident D3-branes and $g_s N$ is large.%
\footnote{That $g_s N$ should be large follows from demanding that the curvature length scale of the supergravity solution of $N$ coincident extremal D3-branes, which is $(N T_{\rm D3}  G)^{1/4} \sim (g_s N)^{1/4} l_s$, should be larger than the string length $l_s$.}
 In this regime the D3-branes backreact on the geometry of the background and we can compute the EM tensor from the supergravity solutions of non-extremal black D3-branes. Doing this, we are essentially employing the recently developed blackfold approach for higher-dimensional black holes \cite{Emparan:2009at,Emparan:2007wm}.

In the blackfold approach \cite{Emparan:2009at,Emparan:2007wm} one finds new black hole solutions by bending flat branes on curved embedding geometries. Indeed, as found in \cite{Emparan:2009at}, a brane with EM tensor $T_{ab}$ and current $J_{abcd}$ with respect to the field strength $F_{(5)}$ would obey the extrinsic equation \eqref{extrinsic} to leading order in the regime where the brane thickness, as defined by its backreacion on the background, is small compared to the length scale of the curvature of the embedding submanifold defined by $X^\mu(\sigma^a)$. This regime is parallel to the regime of the DBI action, $i.e.$ the DBI action is the leading order action in the regime where the length scale of the open string excitations are small as compared to the length scale of the embedding geometry and of the variation of the gauge field on the brane. Higher order terms would for instance take into account derivatives of the $F_{ab}$ gauge field strength.

Below we take the steps to write down the EOMs for thermally excited D3-branes. We first find the appropriate EM tensor in the next subsection,
and remark also on the relation to the extremal case.
Then we write down in Sec.~\ref{sec:solution} the correct EOMs and conservation equations in accordance with the blackfold approach \cite{Emparan:2009at,Emparan:2007wm} and present the solution.
We will also show how the extrinsic equations \eqref{extrinsic} for the thermally excited case can be obtained
from an action, and discuss the relation of the latter to the DBI Hamiltonian.

\subsection{Energy-momentum tensor for black D3-F1 brane bound state}
\label{sec:EMblackbrane}

In this section we consider the EM tensor of D3-branes. The EM tensor \eqref{EMDBI} corresponds to a single D3-brane at zero temperature with a gauge field strength $F_{ab}$ turned on. We work in the regime in which $\gamma_{ab}$ and $F_{ab}$ vary so slowly over the brane that their derivatives can be ignored. The aim of this section is to find the EM tensor of a D3-brane still with the gauge field strength turned on, but now at non-zero temperature and with a large number $N \gg 1$ of parallel stacked D3-branes and with $g_s N$ large.

\subsubsection*{More on DBI case}

Before turning to the non-zero temperature case we take another look at the zero-temperature EM tensor \eqref{EMDBI}. Consider the special case in which the world-volume metric $\gamma_{ab}$ is diagonal, $\gamma_{00}= -1$, and the only non-zero component of the two-form gauge field strength is $F_{01}$. Then the EM tensor \eqref{EMDBI} becomes
\begin{equation}
\label{specialEMDBI}
 T^{00} =
\frac{T_{\rm D3}}{\sqrt{1 - E^2}} \spa T^{11} = -\gamma^{11}\frac{T_{\rm D3}}{\sqrt{1 - E^2}}
\spa T^{ii} = -  \gamma^{ii} T_{\rm D3}\sqrt{1 -
E^2} \ , \ i=2,3
\end{equation}
where we defined
\begin{equation}
E \equiv \frac{2\pi l_s^2 }{\sqrt{\gamma_{11}}} F_{01}
\end{equation}
In the bulk, turning on the electric flux $F_{01}$ on the brane corresponds to having a number of F-strings along the $\sigma^1$ direction of the brane. Call this number of F-strings $k$. From the world-volume point of view we can say that we have $k$ units of electric flux. We can relate $k$ to $F_{01}$ as
\begin{equation}
\label{specialk}
k = - T_{\rm D3} \int_{V_{23}}
d\sigma^2 d\sigma^3 \frac{\partial \sqrt{- \det(\gamma +
2\pi l_s^2 F)}}{\partial F_{01}} =  \frac{T_{\rm D3}}{T_{\rm F1}}  \int_{V_{23}} d\sigma^2
d\sigma^3 \sqrt{\gamma_{22}\gamma_{33}} \frac{E}{\sqrt{1-E^2}}
\end{equation}
If we consider the case in which $\gamma_{ab} $ and $F_{01}$ are constant the above equation can be written as
$k T_{\rm F1} = V_\perp T_{\rm D3} E / \sqrt{1-E^2}$ where $V_\perp$ is the area in the $\sigma^{2,3}$ directions perpendicular to the F-strings. Using this we see that the mass density of the brane can be written as
\begin{equation}
T_{00} =  \sqrt{ T_{\rm D3}^2 + \frac{k^2 T_{\rm F1}^2}{ V_\perp^2} }
\end{equation}
We recognize this as the $1/2$ BPS mass density formula for the D3-F1 brane bound state in the case of a single D3-brane and $k$ F-strings.

\subsubsection*{Energy-momentum tensor for D3-F1 bound state from black brane geometry}

We now turn to obtaining the EM tensor for $N$ D3-branes with an electric field on, corresponding to $k$ units of electric flux, at non-zero temperature. We can obtain this from a black D3-F1 brane bound state geometry assuming that we are in the regime of large $N$ and $g_s N$.

The D3-F1 black brane bound state background has the string frame metric \cite{Harmark:2000wv}
\begin{equation}
\label{D3F1_geom}
ds^2 = D^{-\frac{1}{2}} H^{-\frac{1}{2}} ( - f dt^2 + dx_1^2 ) +
D^{\frac{1}{2}} H^{-\frac{1}{2}} ( dx_2^2 + dx_3^2 ) +
D^{-\frac{1}{2}} H^{\frac{1}{2}} ( f^{-1} dr^2 + r^2 d\Omega_5^2 )
\end{equation}
where
\begin{equation}
\label{fHD}
f = 1 - \frac{r_0^4}{r^4} \spa H = 1 + \frac{r_0^4 \sinh^2
\alpha}{r^4} \spa D^{-1} = \cos^2 \zeta + \sin^2 \zeta H^{-1}
\end{equation}
and with dilaton field $\phi$, Kalb-Ramond field $B_{(2)}$, and two- and four-form Ramond-Ramond gauge fields $C_{(2)}$ and $C_{(4)}$ given by
\begin{equation}
\label{D3F1fields}
\begin{array}{c} \ds
e^{2\phi} = D^{-1} \spa B_{01} = \sin \zeta ( H^{-1} -1 ) \coth \alpha \\[4mm] \ds
C_{23} = \tan
\zeta ( H^{-1} D  - 1) \spa C_{0123} = \cos \zeta D ( H^{-1} - 1 ) \coth \alpha
\end{array}
\end{equation}
We now proceed to read off the EM tensor, the D3-brane and F-string currents, and the thermodynamical parameters as seen by an asymptotic observer. However, we first need to consider how the non-trivial world-volume metric $\gamma_{ab}$ can enter in this. Reading off the EM tensor from \eqref{D3F1_geom}  we find it in the $(t,x^i)$ coordinates for which the world-volume metric is just $-dt^2 + \sum_i (dx^i)^2$. Instead we want the EM tensor with a world-volume metric of the form
\begin{equation}
\gamma_{ab} d\sigma^a d\sigma^b = - d\sigma_0^2 +
\gamma_{11} d\sigma_1^2 +\gamma_{22} d\sigma_2^2 + \gamma_{33}
d\sigma_3^2
\end{equation}
since this is the most general form needed for our computations in this paper, $i.e.$ a diagonal metric without red-shift factor. To transform the resulting EM tensor in the $(t,x^i)$ coordinates to the above world-volume coordinates $\sigma^a$ we can simply make the rescaling $t=\sigma^0$, $x^i = \sqrt{\gamma_{ii}} \sigma^i$, $i=1,2,3$. One could infer here that such a rescaling is problematic since in general the world-volume metric $\gamma_{ab}$ varies according to where we are situated on the brane. However, to the order we are working in we are precisely suppressing the derivative of the metric in the EM tensor, just as they are suppressed in the DBI EM tensor \eqref{EMDBI}. With this in mind we read off the EM tensor in the $\sigma^a$ world-volume coordinates
\begin{equation}
\label{braneEM}
\begin{array}{c} \ds
T^{00} =   \frac{\pi^2}{2} T_{\rm D3}^2 r_0^4 ( 5 + 4
\sinh^2 \alpha ) \spa T^{11} = - \gamma^{11} \frac{\pi^2}{2} T_{\rm
D3}^2 r_0^4 ( 1 + 4 \sinh^2 \alpha )
\\[4mm] \ds
T^{22} = - \gamma^{22} \frac{\pi^2}{2} T_{\rm D3}^2 r_0^4 ( 1 +
4\cos^2\zeta \sinh^2 \alpha ) \spa T^{33} = - \gamma^{33}
\frac{\pi^2}{2} T_{\rm D3}^2 r_0^4 ( 1 + 4\cos^2\zeta \sinh^2 \alpha
)
\end{array}
\end{equation}
using for example \cite{Harmark:2004ch}.
The D3-brane current is
\begin{equation}
\label{D3current}
J^{0123} =  \frac{2\pi^2 T_{\rm D3}^2}{\sqrt{\gamma}} \cos \zeta
r_0^4 \cosh \alpha \sinh \alpha
\end{equation}
The number of D3-branes in the bound state is $N$. Thus, using \eqref{D3current} we find
\begin{equation}
\label{chargequant}
\cos \zeta r_0^4 \cosh \alpha \sinh \alpha = \frac{N}{2 \pi^2 T_{\rm D3}}
\end{equation}
Furthermore, imposing that we have $k$ F-strings gives
\begin{equation}
\label{kformula} \frac{k}{N} =  \frac{T_{\rm D3}}{T_{\rm F1}} \int_{V_{23}}
d\sigma^2 d\sigma^3 \sqrt{\gamma_{22}\gamma_{33}} \tan \zeta
\end{equation}
We also give the thermodynamic quantities associated to
the horizon, that will be used in what follows.  The temperature $T$ and entropy density $\CS$ are
\begin{equation}
\label{temp_entropy}
T =   \frac{1}{\pi r_0 \cosh \alpha} \spa \CS =2\pi^3 T_{\rm D3}^2 V_{(3)}r_0^5 \cosh \alpha
\end{equation}
while the local D3-brane and F-string chemical potentials are
\begin{equation}
\label{chempot}
 \mu_{\rm D3}^{\rm (local)}=  \tanh \alpha \cos \zeta \spa
  \mu_{\rm F1}^{\rm (local)} = \tanh \alpha \sin \zeta
\end{equation}

\subsubsection*{Extremal limit}

As a check we consider here the extremal limit of the EM tensor of the black D3-F1 brane bound state.
We see from \eqref{chargequant} that the extremal limit is to take $r_0 \rightarrow 0$ keeping $\zeta$ and $r_0^4 \cosh \alpha \sinh \alpha$ fixed.
This gives $T=0$ as it should and the EM tensor
\begin{equation}
T^{00} = \frac{ N T_{\rm D3} }{\cos \zeta}
\spa T^{11} = - \gamma^{11} \frac{ N T_{\rm D3} }{\cos
\zeta} \spa
T^{ii} = - \gamma^{ii} N T_{\rm D3}  \cos \zeta\ , \ i=2,3
\end{equation}
along with the formula \eqref{kformula} for $k$. We see that the
above formulas match \eqref{specialEMDBI} and \eqref{specialk}
provided we identify $E = \sin \zeta$ and put $N=1$.

\subsection{Argument for new approach to thermal D-brane probes}
\label{sec:argument}

As reviewed in the introduction, a number of papers in the literature (see \cite{Kiritsis:1999tx,Babington:2003vm,Hartnoll:2006hr,Grignani:2009ua} and later works) have used the classical DBI action to probe finite temperature backgrounds in string theory. In short, this method consists in Wick rotating both the background as well as the classical DBI action, then finding solutions of the EOMs from the classical Euclidean DBI action and finally identifying the radius of the thermal circle of the background with the radius of the thermal circle in the Euclidean DBI action. From the classical solution one can then evaluate physical quantities for the probe such as the energy, entropy and free energy. We dub here this method the "Euclidean DBI probe" method.

We give here a detailed argument for why our thermal D-brane probe, based on the blackfold approach, should give a more accurate way to probe finite temperature backgrounds in string theory than the Euclidean DBI probe method.%
\footnote{The considerations of this section were independently worked out by Roberto Emparan.}

We begin by considering in more detail the Euclidean DBI probe method. For simplicity we stick to a D3-brane (thus in type IIB string theory) but our considerations apply to any D$p$-brane. Consider a type IIB string theory background with metric $g_{\mu\nu}$ and with a RR five-form field strength $F_{(5)}$ turned on (for simplicity we do not consider other RR field strengths and we also assume a constant dilaton). As shown above, the EOMs for the D3-brane DBI action \eqref{DBIaction} take the form
\begin{equation}
\label{LorEOMs}
T_{\rm DBI}^{ab} K_{ab}{}^\rho =  \perp^{\rho \lambda} \frac{1}{4!}
J^{abcd} F_{\lambda abcd}
\end{equation}
where $T_{\rm DBI}^{ab}$ is given by \eqref{EMDBI} and $J^{abcd}$ by \eqref{RRcurrent}.
We now perform a Wick rotation on the background $t = i t_E$ where $t$ is the time coordinate of the background and $t_E$ the corresponding direction in the Euclidean section of the background. Similarly we also perform a Wick rotation for the world-volume time $\tau = i \tau_E$. Then the EOMs \eqref{LorEOMs} become
\begin{equation}
\label{EuclEOMs}
(T_E)_{\rm DBI}^{ab} (K_E)_{ab}{}^\rho =  (\perp_E)^{\rho \lambda} \frac{1}{4!}
(J_E)^{abcd} (F_E)_{\lambda abcd}
\end{equation}
where the subscript E means that it is the Wick rotated quantity where the Wick rotation in both the bulk and on the world-volume are treated as simple linear transformations on the tensors, $e.g.$ $(T_E)_{\rm DBI}^{00} = - T_{\rm DBI}^{00}$ and so on. It is now easily shown that one also obtains the equations \eqref{EuclEOMs} as the EOMs of the Euclidean DBI probe in the Wick rotated background, $i.e.$ by varying the Wick rotated DBI action in the Wick rotated background.
We can conclude from this that there is a one-to-one map between solutions of the EOMs for a DBI probe in a thermal background and the solutions of the EOMs for a Euclidean DBI probe in the Wick-rotated thermal background.

To solve EOMs corresponds to solving certain differential equations under the restriction of certain boundary conditions. The above equivalence between solving the EOMs for a DBI probe in a thermal background and the EOMs for an Euclidean DBI probe in the Wick-rotated thermal background only means that the differential equations are the same, instead the boundary conditions are different. Thus, the equivalence works only locally. Instead there are global differences in being in the Wick-rotated frame or not since one imposes different boundary conditions, in particular regarding the thermal circle direction $t_E$. In the Euclidean probe method one uses the Wick rotated version of the usual static gauge $t_E = \tau_E$ and one imposes that the size of the thermal circle is the same for the probe as for the (Wick rotated) thermal background. Thanks to the static gauge this boundary condition does not enter in the EOMs for the Euclidean probe and is in that sense a global condition on the solution. Therefore, one could think that this global condition is enough to ensure that the probe, using the Euclidean DBI probe method, is in thermal equilibrium with the background. However, we shall now argue that this global condition is not enough since the requirement of thermal equilibrium between the probe and the background changes the EOMs of the probe by changing the EM tensor, and hence the requirement of thermal equilibrium changes the probe not only globally but also locally.

Consider the example of a D3-brane with zero world-volume field strength $F_{ab}=0$ in the $\ads_5\times S^5$ background. In the probe approximation this is described by the DBI action. The EOMs are of the form \eqref{LorEOMs} with $T_{\rm DBI}^{ab} = - T_{\rm D3} \gamma^{ab}$. We notice that the EM tensor locally is Lorentz invariant. This conforms with the fact that the electromagnetic field on the D3-brane is in the vacuum state.

We now turn on the temperature in the background. Thus, the background is either hot AdS space, or an AdS black hole, depending on the temperature, times the $S^5$. The D3-brane should be in thermal equilibrium with the background. This means in particular that the DOFs living on the D3-brane get thermally excited, acquiring the temperature of the background. Among the DOFs are the ones described by the electromagnetic field $F_{ab}$ living on the brane. The quantum excitations of $F_{ab}$ are described by the maximally supersymmetric quantum electrodynamics locally on the brane (since one can see by expanding the DBI action that one locally has Maxwell electrodynamics for small $F_{ab}$ which then is supplemented by the superpartners from the $\CN=4$ supersymmetry). This means that for small temperatures, near extremality, the EM tensor takes the form of the Lorentz invariant piece plus the EM tensor corresponding to a gas of photons and their superpartners. Consider a particular point $q$ on the brane. We can always transform the coordinates locally so that $\gamma_{ab} = \eta_{ab}$ at that point. Then the EM tensor at $q$ takes the form
\begin{equation}
\label{TNE}
T_{ab} = - T_{\rm D3} \eta_{ab} + T^{\rm (NE)}_{ab} \spa T^{\rm (NE)}_{00} = \rho \spa T^{\rm (NE)}_{ii} = p \ , \ i =1,2,3
\end{equation}
where $T^{\rm (NE)}_{ab}$ is the contribution due to the gas of photons and superpartners, having the equation of state $\rho = 3p = \pi^2 T^4 /2$ (the power of $T^4$ follows from the fact that $\CN=4$ supersymmetric quantum electrodynamics is conformally invariant). Thus, we see that the fact that we have local DOFs living on the D3-brane means that the EM tensor is changed once we turn on the temperature. In terms of the EOMs for the probe we see that they are given by \eqref{extrinsic} with the EM tensor \eqref{TNE} (one can easily find this EM tensor for general world-volume metric $\gamma_{ab}$). Therefore, the EOMs are clearly not the same as those of \eqref{LorEOMs} where $T_{\rm DBI}^{ab} = - T_{\rm D3} \gamma^{ab}$. Indeed, the EM tensor \eqref{TNE} is no longer locally Lorentz invariant, which is in agreement with the fact that the brane has an excited gas of photons and superpartners on it.

In conclusion, the above example clearly illustrates that the requirement of thermal equilibrium affects the probe not only globally but also locally in that the EOMs change from those given from the DBI action. This shows that the thermal D-brane probe is not accurately described by the Euclidean DBI probe method.%
\footnote{Note that the difference between the thermal D-brane probe and the Euclidean DBI probe is not due to backreaction. A backreaction would mean that $K_{ab}{}^\rho$ should change. Instead, demanding thermal equilibrium means that the EM tensor changes. Thus, even in the probe approximation the Euclidean DBI probe is not accurate.}

Finally, we note that our description of the thermal D-brane probe is in accordance with the above example. Indeed, if we expand the EM tensor for a non-extremal D3-brane
\begin{equation}
T^{00} =   \frac{\pi^2}{2} T_{\rm D3}^2 r_0^4 ( 5 + 4
\sinh^2 \alpha ) \spa T^{ii} = - \gamma^{ii} \frac{\pi^2}{2} T_{\rm
D3}^2 r_0^4 ( 1 + 4 \sinh^2 \alpha ) \ , \ i=1,2,3
\end{equation}
for small temperatures we get
\begin{equation}
T^{00} = N T_{\rm D3} + \frac{3 \pi^2}{8} N^2 T^4 \spa T^{ii} = \gamma^{ii} ( - N T_{\rm D3} + \frac{\pi^2}{8} N^2 T^4  ) \ , \ i =1,2,3
\end{equation}
We see that this is precisely of the form \eqref{TNE} (with an extra factor of $N$ in the leading part) with the leading part being locally Lorentz invariant and the correction corresponding to a gas of gluons and their superpartners (with the usual factor of $3/4$ since we are at strong coupling~\cite{Gubser:1996de}). This is thus in accordance with our general arguments above. Finally, we note that whereas in the regime of validity of the DBI action the near-extremal correction to the EM tensor \eqref{TNE} should be computed quantum mechanically, we can use classical supergravity to compute the full EM tensor at finite temperature in the large $N$ and large $g_s N$ regime since the classical approximation is reliable in this regime.

\section{Thermal D3-brane configuration with electric flux ending in throat}
\label{sec:solution}

In this section we study the D3-F1 configuration at finite temperature in hot flat space. We
derive the EOMs for the embedding directly and supplement
this with an action-based derivation. We then proceed by solving the equations
and comment on the relation to the DBI solution and Hamiltonian.
Finally, we discuss the regime of validity of the solution.

\subsection{D3-F1 extrinsic blackfold equation \label{sec:BFEOM}}

Our setup is specified by exactly the same type of embedding and boundary
conditions as discussed in Sec.~\ref{sec:DBI} for the extremal case, see Eqs.~\eqref{background}-\eqref{flatindmetric}, \eqref{sigmainfinity}, \eqref{sigma0deriv} and the illustration in Fig.~\ref{fig:setup}.
With the results of the previous section, we are thus ready to compute the D3-F1 extrinsic blackfold equation \eqref{extrinsic},
where the right-hand side is zero since our background is 10D Minkowski space-time so there is no five-form field strength.
For the left-hand side we need to compute the extrinsic curvature tensor \eqref{extrcurv} for the embedding described in \eqref{background}-\eqref{flatindmetric}.
The resulting  non-vanishing components are given by
\begin{equation}
\label{K1}
K_{11}{}^{x_1} = \frac{z''(\sigma)}{1+z'(\sigma)^2} \ ,\ K_{22}{}^{x_1} =\frac{\sigma
z'(\sigma)}{1+z'(\sigma)^2} \ , \ K_{33}{}^{x_1} = \frac{\sigma \sin^2 \theta
z'(\sigma)}{1+z'(\sigma)^2}\ , \
K_{ii}^r = - z'(\sigma) K_{ii}^{x_1}\, ,
\end{equation}
with $i=1,2,3$. Since $K_{ii}^r$ and $K_{ii}^{x_1}$ are proportional, the EOMs
\eqref{extrinsic} for this case becomes simply $T^{ab} K_{ab}{}^{x_1} =0$.
Here $T_{ab}$ is the EM tensor \eqref{braneEM} of the black D3-F1 brane system, where the collective coordinates  $r_0$, $\alpha$, $\zeta$
are now promoted to be functions of $\sigma$.
We then find the following EOM for the D3-F1 blackfold
\begin{equation}
\label{CarterFlat}
\frac{z''}{z' (1+{z'}^2)} = - \frac{2}{\sigma} \frac{1+4\cos^2 \zeta \sinh^2 \alpha}{1+4\sinh^2 \alpha}
\end{equation}
The precise form of the functions  $\zeta(\sigma)$ and $\alpha(\sigma)$ entering
this equation (as well as $r_0 (\sigma)$) follow from a number of constraints on the solution, as we will now discuss.

\subsubsection*{Constraints on solution}

Beyond the EOM \eqref{CarterFlat} we also have further constraints  that follow from charge conservation, of both D3-brane and F-string charge, and constancy of the temperature.
From the point of view of the general blackfold construction discussed in Ref.~\cite{Emparan:2009at,Emparan:2007wm} the
constancy of the temperature and angular velocity is a consequence of stationarity.%
\footnote{More precisely, in the blackfold approach the intrinsic EOMs are
conservation of the world volume EM tensor. For stationary solutions to these
one finds that the temperature and angular velocities (if present) are constant.}
Since in our case the blackfold is static only the temperature is relevant. Furthermore, charge conservation can be seen to follow from
additional EOMs, as will be explained and derived in detail in the forthcoming paper Ref.~\cite{upcoming} which generalizes the construction of neutral blackfolds to charged blackfolds.

We start by considering the F-string charge, given in eq.~\eqref{kformula}, which should be conserved along the $\sigma$ direction. Inserting the induced metric \eqref{flatindmetric}, this
gives that
\begin{equation}
\label{flatkformula}
\kappa = \sigma^2 \tan \zeta
\end{equation}
where the constant $\kappa$  is defined as%
\footnote{Note that, in a slight (but meaningful) abuse of notation
this differs by a factor of $1/N$ from $\kappa$ used in the DBI analysis
in Sec.~\ref{sec:bionsolution}.}
\begin{equation}
\label{defkappa}
\kappa \equiv \frac{k T_{\rm F1}}{4\pi N T_{\rm D3}}
\end{equation}
in terms of the conserved charges $N$ and $k$. Note that we can also write \eqref{flatkformula} as
\begin{equation}
\label{flatkformula2}
\cos \zeta = \frac{1}{\sqrt{1+\frac{\kappa^2}{\sigma^4}}}
\end{equation}
We also need to ensure
the conservation of F-string chemical potential \cite{upcoming} (the quantity defined
as $\int d \sigma \sqrt{\gamma_{11}} \mu_{\rm F1}^{\rm (local)}$ with $\mu_{\rm F1}^{\rm (local)}$ the local F-string chemical potential given in \eqref{chempot})
in the world-volume directions transverse to $\sigma$.
This is automatic since we impose spherical symmetry for the two-spheres
parameterized by the coordinates $\theta,\phi$ transverse to the F-string direction $\sigma$ on the world-volume of the D3-brane.

The equation \eqref{flatkformula2} ensures F-string charge conservation by conserving the number of F-strings $k$ assuming the conservation of the number of D3-branes $N$. The latter is imposed using the formula
given in eq.~\eqref{chargequant}, while  constancy of the
temperature is imposed using Eq.~ \eqref{temp_entropy}. Thus, from Eqs.~\eqref{chargequant}, \eqref{temp_entropy} and \eqref{flatkformula2} we can eliminate $r_0(\sigma)$ and $\zeta(\sigma)$ yielding the constraint
\begin{equation}
\label{othercon3}
 \frac{\sinh \alpha}{\cosh ^3 \alpha} = \frac{\pi^2}{2} \frac{N T^4}{T_{\rm D3}} \sqrt{1 + \frac{\kappa^2}{\sigma^4}}\, .
\end{equation}
Below we solve this equation explicitly for $\cosh \alpha$.

Summarizing, we note that for given $(T,N,k)$ the $\sigma$-dependence
of the two functions $\zeta (\sigma)$ and $\alpha (\sigma) $  is determined by
the two equations \eqref{flatkformula2} and \eqref{othercon3} and.
Given the solution of $\alpha (\sigma)$, one can then compute the thickness $r_0 (\sigma)$ from \eqref{temp_entropy}.

\subsubsection*{Action derivation}

It is also instructive and useful to consider an action derivation of
the EOMs \eqref{CarterFlat}. In fact, as we shall see later, the action derived below helps in finding an analytic solution of the equations
of motion. Moreover, it can be viewed as the action that replaces
the DBI action when thermally exciting the D3-F1 system in the regime
of large number of D3-branes and large $g_s N$.

To write down the action, we use the fact that Refs.~\cite{Emparan:2009at,upcoming} showed that for
stationary blackfolds the extrinsic blackfold equations can be integrated
to an action which is proportional to the Gibbs free energy.%
\footnote{This was done in Ref.~\cite{Emparan:2009at} for neutral stationary blackfolds
and generalized to the charged case in Ref.~\cite{upcoming}. Note also that this implies
that for stationary  blackfolds the extrinsic equations are equivalent to requiring the first law of
thermodynamics.}
We therefore use the thermodynamic action%
\footnote{More properly, the action is $I = \beta \CF$, but since $\beta=1/T$ is
constant we directly use the free energy $\CF$.}
\begin{equation}\label{CarterAction}
 \CF = M-TS
\end{equation}
where $\CF=\CF(T,N,k)$ is the free energy appropriate for the ensemble where the temperature $T$ and number of D3-branes $N$ and F-strings $k$ are fixed.   Here the total mass $M$ and entropy $S$ are found by
integrating the energy density $T^{00}$ in \eqref{braneEM} and the
entropy density $\CS$ in \eqref{temp_entropy} over the D3-brane worldvolume.
\begin{equation}
\label{massentro}
M = \frac{\pi^2}{2}T_{\rm D3}^2 \int dV_{(3)}r_0^4 ( 5 + 4\sinh^2 \alpha ) \spa S =2\pi^3 T_{\rm D3}^2 \int dV_{(3)}r_0^5 \cosh \alpha
\end{equation}
These are thus regarded here as functionals of the embedding function $z(\sigma)$, that give the actual total mass and entropy of the system when evaluated on-shell.
Using \eqref{massentro}  in \eqref{CarterAction} we then find the action functional
\begin{equation}
\CF = \frac{\pi^2}{2}  T_{\rm D3}^2 \int dV_{(3)} r_0^4 ( 1 + 4\sinh^2 \alpha )
\end{equation}
Eliminating $r_0(\sigma)$ using \eqref{temp_entropy} and using the induced metric
\eqref{flatindmetric}  we  get
\begin{equation}
\label{flataction}
\CF = \frac{2 T_{\rm D3}^2}{\pi T^4}   \int_{\sigma_0}^\infty d\sigma \sqrt{1+z'(\sigma)^2}
F(\sigma)
\end{equation}
where we introduced the function
\begin{equation}
\label{Ffunction}
F(\sigma) = \sigma^2  \frac{ 4 \cosh^2 \alpha -3 }{\cosh^4 \alpha}
\end{equation}
Note that we integrate from $\sigma_0$ to infinity in \eqref{flataction} according to the boundary conditions discussed above.

 The function $F(\sigma)$ defined in \eqref{Ffunction} is a specific function of $\sigma$ for given $T$, $N$ and $k$, as seen from \eqref{othercon3}.
 This means that when we vary the action \eqref{flataction} with respect to $z(\sigma)$ the function $F(\sigma)$ does not vary.
Performing this variation in the action \eqref{flataction} we then
find that the EOMs take the form%
\footnote{This is computed as
$
\delta ( \sqrt{1+{z'}^2} F ) = \frac{z'F}{\sqrt{1+{z'}^2}} \delta z'$
and adding a total derivative to the last formula.}
\begin{equation}
\label{theeom}
\left( \frac{z'(\sigma) F(\sigma)}{\sqrt{1+z'(\sigma)^2 }} \right)' = 0
\end{equation}
which we will use below when solving the system.
It is noteworthy that the EOM \eqref{theeom} is exactly
as that of \eqref{theeom0} for the DBI case with $F_{\rm DBI} $ replaced by $F$ in \eqref{Ffunction}. The same holds in fact when comparing the free energy
\eqref{flataction} and the DBI Hamiltonian \eqref{CMham}. In Sec.~\ref{sec:extrbranch} we will further comment on their relation.

As a check on the action approach, we now show that \eqref{theeom} obtained from the thermodynamic
 action is consistent with the D3-F1 extrinsic blackfold equation \eqref{CarterFlat}.
From \eqref{theeom} we find
\begin{equation}
\frac{z''}{z' ( 1 + {z}'^2 )} = - \frac{F'}{F}
\end{equation}
Computing $F'(\sigma)$ from \eqref{Ffunction} and using \eqref{flatkformula2} and \eqref{othercon3} to eliminate $\alpha'(\sigma)$ and $\kappa$, we get
\begin{equation}
\label{dum3}
F' =  \frac{2\sigma}{\cosh ^4 \alpha} \Big[ 1 + 4 \cos^2 \zeta \sinh^2 \alpha   \Big]
\end{equation}
Using this with \eqref{Ffunction} we indeed obtain \eqref{CarterFlat} from \eqref{theeom}. Note that the equivalence of these two equations implies that the blackfold EOMs are equivalent to requiring the first law of thermodynamics, since the latter follows from extremizing the free energy functional.

\subsection{Solution and bounds}

We now proceed solving the EOM \eqref{theeom} subject to the constraints
\eqref{flatkformula2} and \eqref{othercon3}. The latter imply a bound on the temperature as well
as the world-volume coordinate  $\sigma$ parameterizing the size of the two-sphere, which we will
first discuss. Then we turn to the explicit solution of the constraint \eqref{othercon3}  in terms of the constants
 $(T,N,k)$  and subsequently present the solution of the EOMs for the profile $z(\sigma)$.

\subsubsection*{Bounds on the temperature and $\sigma$}

Considering the left hand side  of the constraint \eqref{othercon3} it is easy to check that $\sinh \alpha / \cosh^3 \alpha$ is bounded from above, with a maximal value $2\sqrt{3}/9$ for $\cosh^2 \alpha = 3/2$.
Using \eqref{flatkformula2} we thus get an upper bound for the temperature
\begin{equation}
\label{Tupperbound}
T^4 \leq T^4_{\rm bnd} \cos \zeta
\end{equation}
for given $(N,k)$ and $\sigma$, where we defined the temperature
\begin{equation}
\label{Tbnd}
T_{\rm bnd} \equiv \left( \frac{4 \sqrt{3} T_{\rm D3} }{9 \pi^2 N}  \right)^{\frac{1}{4}}
\end{equation}
From \eqref{Tupperbound} it is obvious that we furthermore have the weaker upper bound $T \leq T_{\rm bnd}$ which only depends on $N$ and thus not on $k$ and $\sigma$. The temperature $T_{\rm bnd}$ is the maximal temperature for $N$ coincident non-extremal D3-branes, which can never be reached in the presence of non-zero F-string charge.
 For future convenience we furthermore define the rescaled temperature
\begin{equation}
\label{Tbar}
\bar{T} \equiv \frac{T}{T_{\rm bnd}}
\end{equation}
which will simplify expressions below. Note from \eqref{Tupperbound} that we have $\bar{T} \leq 1$.

Since $\sigma$ takes values in $[\sigma_0,\infty)$ we see that
$\cos \zeta$ in \eqref{flatkformula2} is minimized for $\sigma = \sigma_0$. It then follows that the upperbound \eqref{Tupperbound}  can be stated more accurately
as
\begin{equation}
\label{Tupperbound2}
T \leq T_{\rm bnd}  \left( 1 + \frac{\kappa^2}{\sigma_0^4} \right)^{-\frac{1}{8}}
\end{equation}
for given $N$, $k$ and $\sigma_0$ of the brane profile $z (\sigma)$.
Note however that generically this bound cannot be saturated for a given $N$, $k$ and $\sigma_0$, so that
generically the maximal temperature is lower than this bound.
Finally, we remark that using \eqref{flatkformula2} the upper bound on the temperature \eqref{Tupperbound} can be turned into a lower bound on  $\sigma$
\begin{equation}
\label{sigmamin}
\sigma \geq \sigma_{\rm min} \equiv \sqrt{\kappa} \left( \bar{T}^{-8} - 1 \right)^{-\frac{1}{4}}
\end{equation}
for given $(T,N,k)$.

\subsubsection*{Solving $\cosh \alpha$ in terms of $\sigma$}

We now present the explicit solution of the constraint \eqref{othercon3}.
Using the upper bound \eqref{Tupperbound} on the temperature along with \eqref{flatkformula2} we can define the angle $\delta (\sigma) $ by
\begin{equation}
\label{cosdelta}
\cos \delta (\sigma) \equiv \bar{T}^4 \sqrt{1+ \frac{\kappa^2}{\sigma^4}}
\end{equation}
where we restrict the angle to be in the interval $0 \leq \delta(\sigma) \leq \pi/2$.
In terms of this angle the constraint \eqref{othercon3} can be written as
\begin{equation}
\label{thirdordereq}
\frac{4 \cos^2 \delta}{27}   \cosh^6 \alpha - \cosh^2 \alpha + 1 = 0
\end{equation}
which is a cubic equation in $\cosh^2 \alpha$. Thus, it has three independent solutions for $\cosh^2 \alpha$ in terms of $\delta$. The first solution is
\begin{equation}
\label{mainbranch}
\cosh^2 \alpha = \frac{3}{2} \frac{\cos \frac{\delta}{3} + \sqrt{3} \sin \frac{\delta}{3} } {\cos \delta}
\end{equation}
As $\delta$ goes from $0$ to $\pi/2$, the right hand side increases monotonically from $3/2$ to infinity. Substituting $\delta \rightarrow -\delta$ in \eqref{mainbranch} we find the second solution
\begin{equation}
\label{branch2}
\cosh^2 \alpha = \frac{3}{2} \frac{ \cos \frac{\delta}{3} -\sqrt{3} \sin \frac{\delta}{3} }{\cos \delta}
\end{equation}
Here the right hand side decreases monotonically from $3/2$ to $1$ as $\delta$ goes from $0$ to $\pi/2$.
Finally, by substituting $\delta \rightarrow \delta - 2\pi$ in \eqref{mainbranch} we find the third solution
\begin{equation}
\cosh^2 \alpha = - \frac{3}{2} \frac{ \cos \frac{\delta+\pi}{3} +\sqrt{3} \sin \frac{\delta+\pi}{3} }{\cos \delta}
\end{equation}
This solution can be immediately discarded since it decreases from $-3$ to $-\infty$ as $\delta$ goes from $0$ to $\pi/2$, and $\alpha$ has to be real.

Turning to the two solutions \eqref{mainbranch} and \eqref{branch2} we note that both of them respect that $\cosh^2 \alpha \geq 1$. In the extremal limit one takes $\alpha \rightarrow \infty$. Therefore, the solution branch connected to the extremal solution is the first solution \eqref{mainbranch}. For this branch the energy density
($T^{00}$ in \eqref{braneEM}) for each value of $\sigma$ increases as the temperature $T$ increases holding $(N,k)$ fixed. Thus, this is the thermodynamically stable branch with positive heat capacity. Instead in the second solution branch \eqref{branch2} the energy density at each $\sigma$ decreases with increasing temperature, thus resulting in a negative heat capacity. This branch is connected to the neutral 3-brane with $\alpha=0$. The two branches meet at the point $\cosh^2 \alpha = 3/2$.

\subsubsection*{The solution}

Finally, we turn to the solution of the EOM \eqref{theeom}. Imposing the boundary condition
\eqref{sigma0deriv}  we find that the solution takes form%
\footnote{Notice that it follows from the expression \eqref{zprimesolution} that the derivatives $d^n \sigma / dz^n$ of the inverse function $\sigma( z)$ vanish for odd $n$ at $z = z(\sigma_0)$. This is a necessary requisite for the smoothness of the wormhole solution of Section \ref{sec:braneseparation}.}
\begin{equation}
\label{zprimesolution}
- z'(\sigma) = \left( \frac{F(\sigma)^2}{F(\sigma_0)^2} - 1 \right)^{-\frac{1}{2}}
\end{equation}
Imposing the boundary condition \eqref{sigmainfinity} we find
\begin{equation}
\label{thesolution}
z(\sigma) = \int_{\sigma}^\infty d{\sigma'} \left( \frac{F({\sigma'})^2}{F(\sigma_0)^2} - 1 \right)^{-\frac{1}{2}} \spa F(\sigma) = \sigma^2  \frac{ 4 \cosh^2 \alpha -3 }{\cosh^4 \alpha}
\end{equation}
with $\sigma \geq \sigma_0$ and where we repeated the definition of $F(\sigma)$ from \eqref{Ffunction} given in terms of $\cosh^2 \alpha$. Furthermore, $\cosh^2 \alpha (\sigma)$  is given by one of the two branches \eqref{mainbranch} and \eqref{branch2}.
The result \eqref{thesolution} describing the heated up BIon solution
 is one of the central results of this paper. Just as the BIon solution,
it is obtained in a probe approximation, which is discussed in more
detail in Sec.~\ref{sec:validity}.

It should be emphasized that for each of the two branches we have an explicit expression for the derivative of the brane profile $z'(\sigma)$
so that the mass and entropy for each solution branch
 can be obtained explicitly  from the integrands in \eqref{massentro}, \eqref{flataction}. This gives
\begin{equation}
\label{Msolution}
M =\frac{2 T_{\rm D3}^2}{\pi T^4} \int_{\sigma_0}^\infty d \sigma
        \frac{F(\sigma)}{ \sqrt{F^2(\sigma)-F^2(\sigma_0)} } \,\sigma^2 \frac{4 \cosh^2 \alpha+ 1}{\cosh^4 \alpha} \spa
\end{equation}
\begin{equation}
\label{Ssolution}
S =   \frac{2 T_{\rm D3}^2}{\pi T^5} \int_{\sigma_0}^\infty d \sigma
        \frac{F(\sigma)}{\sqrt{F^2(\sigma)-F^2(\sigma_0)} } \,\sigma^2 \frac{4}{\cosh^4 \alpha}
\end{equation}
We also give the integrated chemical potentials \cite{upcoming} that follow from \eqref{chempot}%
\footnote{We redefine these for convenience by an extra factor
of the D3-brane and F-string tension respectively.}

\begin{equation}
\label{mu3solution}
\mu_{\rm D3} = 4 \pi T_{\rm D3}  \int_{\sigma_0}^\infty d \sigma
        \frac{F(\sigma)}{\sqrt{F^2(\sigma)-F^2(\sigma_0)} } \,\sigma^2
         \tanh \alpha \cos \zeta
\end{equation}
\begin{equation}
\label{mu1solution}
\mu_{\rm F1} = T_{\rm F1} \int_{\sigma_0}^\infty d \sigma
        \frac{F(\sigma)}{\sqrt{F^2(\sigma)-F^2(\sigma_0)} } \,
         \tanh \alpha \sin \zeta
\end{equation}
These satisfy the first law of thermodynamics and Smarr relation
\begin{equation}
d M = T d S + \mu_{\rm D3} d N +  \mu_{\rm F1} d k \spa
4 (M- \mu_{\rm D3} N - \mu_{\rm F1} k) = 5 TS
\end{equation}
The integrations in \eqref{Msolution}-\eqref{mu1solution} can be performed
numerically or analytically in certain limits.

The solution presented above gives the profile of a configuration of $N$ coincident infinitely extended D3-branes with $k$ units of F-string charge, ending in a throat with minimal radius $\sigma_0$, at temperature $T$. The configuration is sketched in Fig.~\ref{fig:wh_bottom}. As explained
in Sec.~\ref{sec:DBI} we can construct a corresponding wormhole solution by attaching a mirror solution as illustrated in Fig.~\ref{fig:wh}. This will be discussed in Sec.~\ref{sec:braneseparation}.
In the next subsection we examine two limits of the branch \eqref{mainbranch} connected
to the extremal configuration, which will be used in Sec.~\ref{sec:braneseparation}.
App.~\ref{sec:neutbranch} presents some features of the branch \eqref{branch2} connected
to the neutral 3-brane, which will not be further discussed in the main text.

\subsection{Analysis of branch connected to extremal configuration}
\label{sec:extrbranch}

Our main focus in this paper is the solution \eqref{thesolution} for the branch \eqref{mainbranch} connected to the extremal configuration, which we start analyzing here by considering two physically relevant limits. We then go on to showing the equivalence of our thermal action (free energy) \eqref{flataction} in the zero temperature limit to the DBI Hamiltonian.

One interesting limit is the limit of small temperature, which enables to
see in a small temperature expansion the effect of heating up the BIon as
compared to the extremal case. The temperature enters the solution through
the function $\alpha (\sigma)$, which is given by \eqref{mainbranch}
for the branch connected to the extremal configuration. From \eqref{cosdelta}
it is seen that for small temperatures $\delta$ is close to $\pi/2$,
so that
\begin{equation}
\cosh^2 \alpha = \frac{3\sqrt{3}}{2\cos \delta} - \frac{1}{2} - \frac{\sqrt{3}}{12} \cos \delta - \frac{2}{27} \cos^2 \delta + \CO (\cos^3 \delta )
\end{equation}
This gives for the function $F(\sigma)$ defined in \eqref{Ffunction} the expansion
\begin{equation}
\label{FsmallT1}
\frac{F}{\sigma^2} = \frac{8\sqrt{3}}{9} \cos \delta - \frac{4}{27} \cos^2 \delta - \frac{4\sqrt{3}}{243} \cos^3 \delta + \CO( \cos^4 \delta )
\end{equation}
Keeping the first two terms in the expansion we then find from
the solution \eqref{zprimesolution} the result
\begin{equation}
\label{zsmallT}
-z' (\sigma) = \sqrt{ \frac{\kappa^2 + \sigma_0^4}{\sigma^4 - \sigma_0^4 } } \left[ 1 + \frac{\sqrt{3}}{18} \bar{T}^4 \frac{\kappa^2+\sigma^4 }{\sigma^4 - \sigma_0^4 } \left( \sqrt{1+\frac{\kappa^2}{\sigma^4}} - \sqrt{1+\frac{\kappa^2}{\sigma_0^4}} \right)
+\mathcal{O}(\bar{T}^{8})
 \right]
\end{equation}
Taking the zero temperature limit, we see that the first term agrees
with the BIon solution \eqref{Xprime}. The second term  describes the leading order
effect due to heating up the BIon. Note that the factor in front of $\bar{T}^4$ does not blow up as $\sigma\rightarrow \sigma_0$. This expression is a good approximation for all $\sigma \geq \sigma_0$ when $(T,N,k)$ and $\sigma_0$ are given such that  $\bar{T}^4 \sqrt{1+ \kappa^2 /\sigma_0^4} \ll 1$.

Another interesting limit is $\sigma/\sqrt{\kappa} \gg 1$. This limit
describes the profile for large $\sigma$, $i.e.$ in the region close to
the flat D3-brane at infinite $\sigma$. In this limit we
have to leading order $\cos \delta = \bar{T}^4$.
Therefore, from \eqref{mainbranch} we see $\cosh^2 \alpha$ is a function of $\bar{T}$ only. For large $\sigma/\sqrt{\kappa}$ we thus have
\begin{equation}
\label{Flargesigma}
F(\sigma) = \sigma^2 g( \bar{T} ) + \CO( \kappa /\sigma^2 )
\end{equation}
where $g(\bar{T})$ is a function that increases from $0$ to $4/3$ as $\bar{T}$ goes from $0$ to $1$. For sufficiently large $\sigma$ we find therefore
from \eqref{thesolution} the behavior
\begin{equation}
z(\sigma) = \frac{F(\sigma_0)}{g(\bar{T}) } \frac{1}{\sigma} + \CO( \sigma^{-5} )
\end{equation}
If we require $\sigma/\sqrt{\kappa} \gg1$ for all $\sigma$, the limit
corresponds to $\sigma_0 /\sqrt{\kappa} \gg 1$, which can be regarded
as the limit in which the F-string charge is taken to be small.
In this case it follows from \eqref{Flargesigma} that $F(\sigma) / F(\sigma_0) \simeq \sigma^2 / \sigma_0^2$ and hence
\begin{equation}
\label{zprimelargesigma}
-z'(\sigma) \simeq \frac{\sigma_0^2 }{\sqrt{\sigma^4 -\sigma_0^4}}
\end{equation}
which agrees with \eqref{Xprime} in the limit $\sigma_0 /\sqrt{\kappa} \gg 1$.

\subsubsection*{Comparison with DBI Hamiltonian}

Finally, we point out the relation of the DBI action with our thermal D3-F1 brane action (free energy) in \eqref{flataction}.
To this end we compute the $T\rightarrow 0$ limit of our action. In particular, using the small temperature expansion \eqref{FsmallT1} for the branch connected to the extremal configuration
along with the expression \eqref{cosdelta} for $\cos \delta$ and the
definition \eqref{Tbnd} we find by inserting in the action \eqref{flataction}
that
\begin{equation}
\label{equivalence}
\lim_{T \rightarrow 0} {\cal{F}} = H_{\rm DBI} \vert_{T_{D3} \rightarrow N T_{D3}}
\end{equation}
This shows that our thermodynamic action \eqref{CarterAction} can be viewed as the thermalization of the DBI Hamiltonian \eqref{CMham}.
The replacement $T_{D3} \rightarrow N T_{D3}$ induces an extra
factor of $N$ in front of the DBI Hamiltonian \eqref{CMham}, and
at the same time includes an extra factor of $1/N$ in $\kappa$, yielding
 the one defined in  \eqref{defkappa}. It is satisfying to
recover these multiplicative factors of $N$ since the DBI action is valid for a single D3-brane only and our thermal system
contains instead $N$ D3-branes. Moreover, this relation implies that the systems are properly connected also off-shell
and makes manifest that the extremal limit of the branch \eqref{mainbranch} of our solution gives the BIon solution reviewed in Section \ref{sec:bionsolution}.

\subsection{Validity of the probe approximation \label{sec:validity}}

In the blackfold approach used to obtain the EOM
\eqref{CarterFlat}, the D3-F1 system is considered in the probe approximation. We therefore need
to determine  the conditions for which this is a valid approximation. If we denote the transverse size scale of the
D3-F1 geometry by $r_{\rm s}$, there are two conditions that have to be met across the embedded surface
in order for gravitational backreaction to be negligible:
\begin{itemize}
\item The size scale should be much less than the $S^2 $ radius  $\sigma $ of the induced geometry:  $r_{\rm s} \ll \sigma $
\item The size scale should be much less than the length scale $L_{\rm curv}$ of the extrinsic curvature of the induced geometry: $r_{\rm s}  \ll L_{\rm curv}$.
\end{itemize}

The transverse size scale $r_s$ of the D3-F1 geometry \eqref{D3F1_geom} is determined by the harmonic functions $f$, $H$ and $D$ in \eqref{fHD}.  The three corresponding scales are $r_0$, $r_0 \sinh^{1/2} \alpha$ and $r_0 (\sin \zeta \sinh \alpha)^{1/2}$ respectively. Since $\sin^2 \zeta \leq 1$, we only need to consider the first two scales and we let $r_s$ be the largest of the two.
For the main solution branch \eqref{mainbranch} connected to the extremal configuration the largest scale
is then $r_{\rm s}^4 = r_0^4 \sinh^2 \alpha$, since $ 1/2 \leq \sinh^2 \alpha < \infty$.
Moreover, it is in fact sufficient
to consider the charge radius $r_{\rm c}^4 = r_0^4 \sinh \alpha \cosh \alpha$ instead. Near extremality this
is obviously true, but, more generally we have on this branch $ 1/3 \leq \tanh^2 \alpha \leq 1$, so that $r_{\rm c}$ differs by a factor of order 1 from $r_{\rm s}$.
For the other branch \eqref{branch2} connected to the neutral 3-brane, where we have $0 \leq \sinh^2 \alpha \leq 1/2$, similar reasoning implies that it is sufficient
to consider the radius $r_{\rm s} = r_0$.
Focusing on the main branch, the above conditions can thus be written as
\begin{equation}
\label{cond1}
r_{\rm c}(\sigma) \ll \sigma \spa  r_{\rm c} (\sigma) \ll L_{\rm curv} (\sigma)
\end{equation}
where, using \eqref{chargequant} and \eqref{flatkformula2}, the charge radius is found to be
\begin{equation}
\label{chargeradius}
r_{\rm c}^4 \sim \frac{N}{T_{\rm D3}} \sqrt{ 1 +\frac{\kappa^2}{\sigma^4} }
\end{equation}
Considering the first condition $r_c(\sigma) \ll \sigma$ in \eqref{cond1} we see that $r_c(\sigma)$ is largest for $\sigma=\sigma_0$ thus to satisfy $r_c(\sigma) \ll \sigma$ for all $\sigma \geq \sigma_0$ it is sufficient to demand $r_c (\sigma_0 ) \ll \sigma_0$.

To understand the second condition $r_c(\sigma)\ll \sigma$ in \eqref{cond1} we first need to determine $L_{\rm curv}(\sigma)$.
For this we need to compute the mean curvature vector $K^{\rho} = \gamma^{ab} K_{ab}^{\rho}$.
The curvature size is then given by $L_{\rm curv} = |K|^{-1}$ with $K = K^{\rho } n_{\rho}$, where
$n_\rho$ is the unit normal vector of the embedding surface of the three-brane.
In the case at hand we have
\begin{equation}
n_{\rho} = \frac{1}{\sqrt{1+z'(\sigma)^2}}  (0,-z'(\sigma),0,0,1,\vec{0}_5)
\end{equation}
in the coordinates $(t,r,\theta,\phi,x_i)$ used in \eqref{background}. To see this note that the tangent
vector to the 3-brane is $V^{\rho} = \frac{1}{\sqrt{1+z'(\sigma)^2}}  (0,1,0,0,z'(\sigma),\vec{0}_5)$,
so that $n_\rho V^\rho =0$. Using the second fundamental tensor computed in \eqref{K1} we then find
after some algebra
\begin{equation}
\label{Kexp}
K = \frac{1}{(1+z'{}^2)^{3/2}} \left[ z''  + 2 \frac{z'}{\sigma} ( 1 + z'{}^2) \right]
= \frac{F_0}{F} \left[ \frac{F'}{F} - \frac{2}{\sigma} \right]
= - \frac{2}{\sigma} \frac{F_0}{F} \frac{ 4  \sinh^2 \alpha}{1 + 4 \sinh^2 \alpha} \sin^2 \zeta
\end{equation}
 where $F_0 \equiv F(\sigma_0)$. To examine this further, we note that
 for the main branch we have $ \sinh^2 \alpha \geq 1/2$ so that
 the the third factor in \eqref{Kexp}  is of order one. We have furthermore checked that the extrinsic curvature \eqref{Kexp} reaches its maximum $K_{\rm max}$ in the region $\sigma \sim \sigma_0$ close to the end of the throat. We thus get
 \begin{equation}
 K_{\rm max} \sim - \frac{1}{\sigma_0} \sin^2 \zeta
   \end{equation}
Hence the second condition in \eqref{cond1} requires to be in the regime where  $r_c (\sigma_0) \ll \sigma_0/\sin^2 \zeta$. However, this is already ensured by the stronger condition $r_c (\sigma_0) \ll \sigma_0$ coming from the first condition
in \eqref{cond1}. Thus, in conclusion, it is sufficient to impose the condition 
\begin{equation}
\label{cond2}
r_c (\sigma_0) \ll \sigma_0
\end{equation}
to ensure the validity of the probe approximation. 
To get a better understanding of the condition \eqref{cond2}, we now consider it in the regime where $\sigma_0/\sqrt{\kappa} $ is very small. Using the definition of  $\kappa$ in \eqref{defkappa} the condition becomes $\sigma_0^3 \gg \sqrt{k} g_s l_s$. It is interesting to note that the $N$-dependence has canceled out in this condition. It is intuitively clear
that the larger the number of F-strings, the greater the minimum radius should be in order to neglect backreaction.
Moreover, for sufficiently weak string coupling it is always possible to be in the correct range.

\section{Separation between branes and anti-branes in wormhole solution}
\label{sec:braneseparation}


\subsection{Brane-antibrane wormhole solution \label{sec:wormholesolution}}

In Section \ref{sec:solution} we found a configuration where $N$ coincident thermally excited D3-branes have an electric flux such that the equipotential surfaces of the electric field are on two-spheres. The shape of the configuration is given by the function $z(\sigma)$ in Eq.~\eqref{thesolution}. As $\sigma$ ranges from $\infty$ to $\sigma_0$ the bulk coordinate $z(\sigma)$ ranges from zero to $z(\sigma_0)$ where the tangent of the brane is orthogonal to the tangent of the brane at $\sigma=\infty$. Thus, we have $k$ electric flux lines all pointing towards a center, and as $\sigma$ decreases the density of the electric flux on the equipotential two-spheres increases. However, rather than having a central singularity as in the linear theory of Maxwell electrodynamics, the non-linear nature of the DBI theory on the D3-brane gives instead a bending of the brane preventing us from reaching the point where the singularity should have been. The spherically symmetric electric flux thus causes the bending of the D3-brane in the bulk creating a throat on the brane ending at $r=\sigma_0$ and $z=z(\sigma_0)=\Delta/2$ as illustrated in Figs.~\ref{fig:setup} and \ref{fig:wh_bottom}. As stated in Section \ref{sec:solution} we should distinguish between solutions $z(\sigma_0)=\infty$ and $z(\sigma_0) <\infty$. For $z(\sigma_0)=\infty$ we have an infinite spike. Instead, for $z(\sigma_0) <\infty$ the solution as it is violates the conservation of D3-brane and F-string charge. To remedy this one can attach a mirror of the solution, reflected in $x_1$ around $x_1=z(\sigma_0)= \Delta/2$, as illustrated in Fig.~\ref{fig:wh}. This we call a brane-antibrane wormhole solution. 

In this section we shall see that it is not possible to find finite-temperature solutions with $z(\sigma_0)=\infty$, $i.e.$ the infinite spike, unlike in the zero temperature case (see Section \ref{sec:bionsolution}). 
For this reason, we focus here on the brane-antibrane
wormhole solution. However, a different way of generalizing the extremal infinite spike solution to finite temperature will be considered in \cite{forthcoming}.

In more detail, the brane-antibrane wormhole solution has $N$ coincident D3-branes extending to infinity at $x_1 = 0$, and $N$ anti D3-branes ($i.e.$ oppositely charged) extending to infinity at $x_1 = \Delta = 2z(\sigma_0)$. Then centered around $r=0$ we have a wormhole where for each constant $x_1$ slice you have a two-sphere where a flux of $k$ F-strings is going through in the positive $x_1$ direction. At $x_1=z(\sigma_0)$ the two-sphere has the minimal radius $r=\sigma_0$. The solution has altogether four parameters: The D3-brane charge $N$, the F-string charge $k$, the temperature $T$ and the minimal radius $\sigma_0$. It is important to note here that the extensive quantities given in 
\eqref{Msolution}-\eqref{mu1solution} should be multiplied by a factor of 2 since we add a mirror of the solution.

In this section we consider the separation distance $\Delta = 2z(\sigma_0)$ between the $N$ D3-branes and $N$ anti D3-branes for a given brane-antibrane wormhole configuration defined by the four parameters $N$, $k$, $T$ and $\sigma_0$. We consider only the branch \eqref{mainbranch} connected to the extremal solution. Thus, we have from \eqref{thesolution}
\begin{equation}
\label{braneseparation}
\Delta = 2 \int_{\sigma_0}^\infty d\sigma' \left( \frac{F(\sigma')^2}{F(\sigma_0)^2} - 1 \right)^{-\frac{1}{2}}
\end{equation}
with $F(\sigma)$ given by \eqref{Ffunction}, \eqref{mainbranch} and \eqref{cosdelta}. From considering  \eqref{Ffunction}, \eqref{mainbranch} and \eqref{cosdelta} we see that $F(\sigma)$ depends on $N$, $k$ and $T$ only through the variables $\kappa$ and $\bar{T}$ defined in \eqref{defkappa} and \eqref{Tbar}. Therefore, the separation distance is a function with dependence $\Delta = \Delta (\bar{T} , \sigma_0 , \kappa)$. However, changing $\kappa$ corresponds to a uniform scaling of the system. Indeed, it is easy to show that $\Delta(\bar{T} , \sigma_0 , \kappa)$ has the scaling property
\begin{equation}
\Delta ( \bar{T}  , \sigma_0 , \kappa ) = \sqrt{\kappa} \, \Delta \Big( \bar{T}, \frac{\sigma_0}{\sqrt{\kappa}} , 1   \Big)
\end{equation}
Thus, it is enough to find $\Delta$ for $\kappa=1$. In detail, given $\Delta$ for a certain $\sigma_0$ and $\bar{T}$ with $\kappa=1$ the general $\kappa$ configuration is found by rescaling $\Delta \rightarrow \sqrt{\kappa} \Delta$ and $\sigma_0 \rightarrow \sqrt{\kappa}\sigma_0$ while keeping $\bar{T}$ fixed.

\subsection{Diagrams for separation distance $\Delta$ versus minimal radius $\sigma_0$}

In the following we examine the behavior of the separation distance $\Delta (\bar{T}, \sigma_0, \kappa=1) \equiv \Delta(\bar{T}, \sigma_0)$. We set $\kappa=1$ in the rest of this section since, as argued above, one can always reinstate $\kappa$ by a trivial rescaling.

\begin{figure}[h!]
\centerline{\includegraphics[scale=0.9]{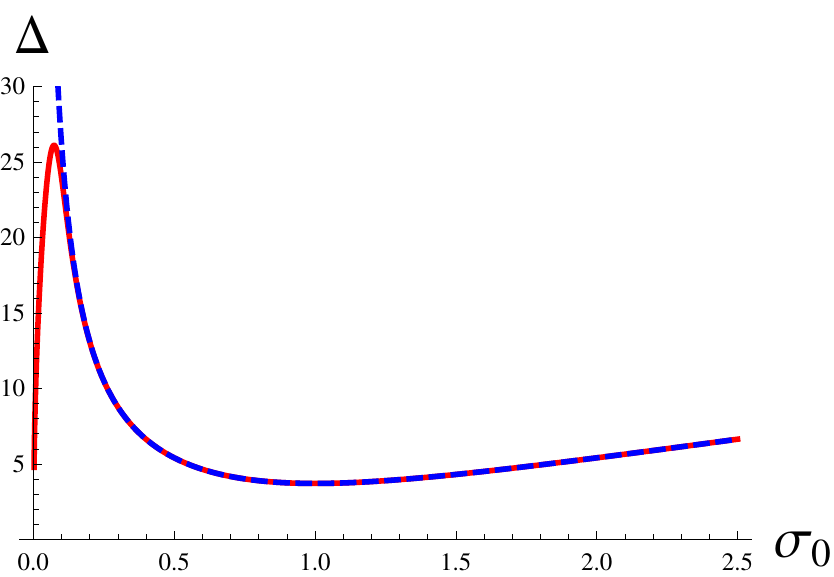} \includegraphics[scale=0.9]{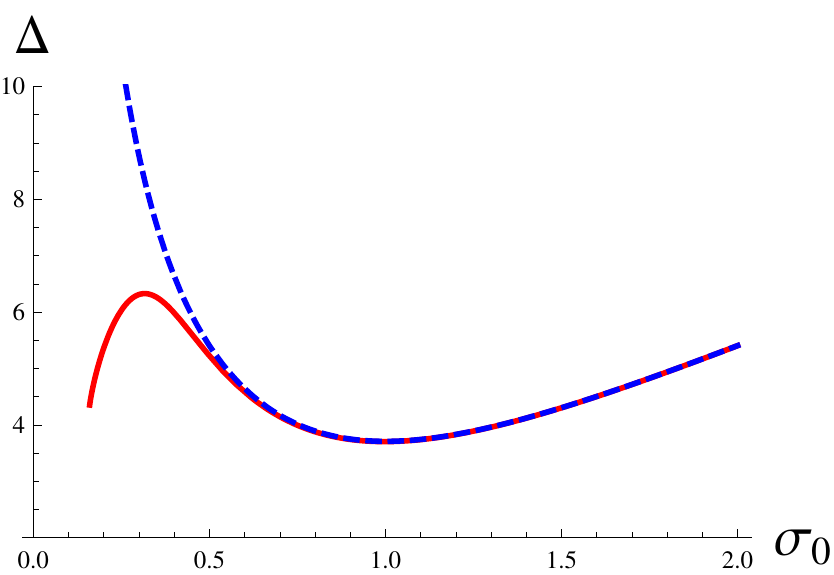}}
\caption{\small On the figures the solid red line is $\Delta$ versus $\sigma_0$ either for  $\bar{T}=0.05$ (left figure) or $\bar{T}=0.4$ (right figure) while the blue dashed line corresponds to $\bar{T}=0$. We have set $\kappa=1$.}
\label{DeltaPlot41}
\end{figure}

\begin{figure}[h!]
\centerline{\includegraphics[scale=0.9]{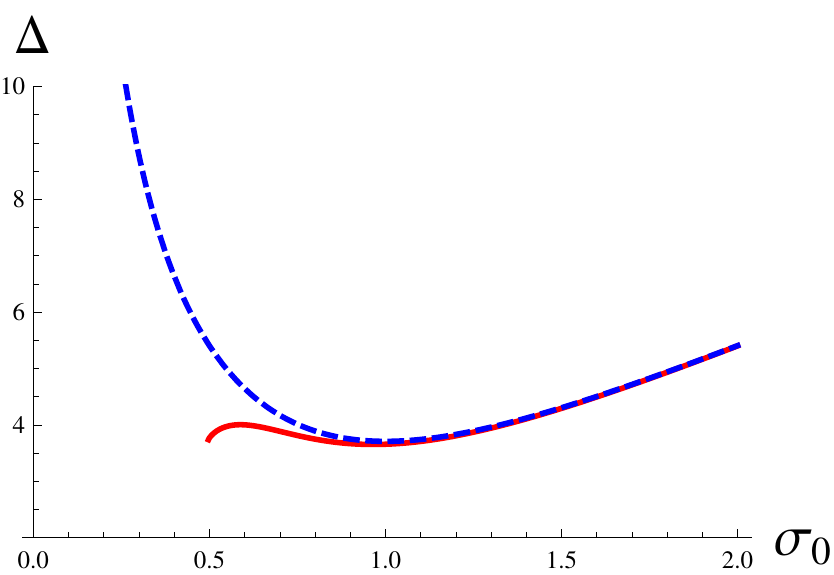} \includegraphics[scale=0.9]{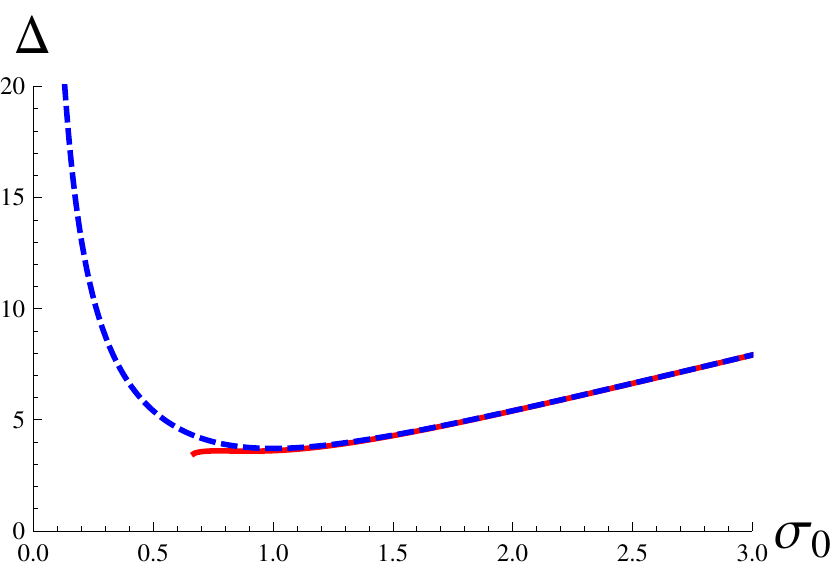}}
\caption{\small On the figures the solid red line is $\Delta$ versus $\sigma_0$ either for  $\bar{T}=0.7$ (left figure) or $\bar{T}=0.8$ (right figure) while the blue dashed line corresponds to $\bar{T}=0$. We have set $\kappa=1$.}
\label{DeltaPlot25}
\end{figure}

We shall examine the separation distance $\Delta (\bar{T}, \sigma_0)$ by using both numerical and analytical methods.
We first evaluate $\Delta(\bar{T}, \sigma_0)$  numerically. The behavior of $\Delta$ versus $\sigma_0$ for four different values of $\bar{T}$ is shown in Figures~\ref{DeltaPlot41} and \ref{DeltaPlot25}, where we compare it with the case of zero temperature (Fig.\ref{DeltaCM}). We have chosen the values $\bar{T} = 0.05, 0.4, 0.7, 0.8$. Note that $0 \leq \bar{T} \leq 1$.

\begin{figure}[ht]
\centering
\includegraphics[scale=1.1]{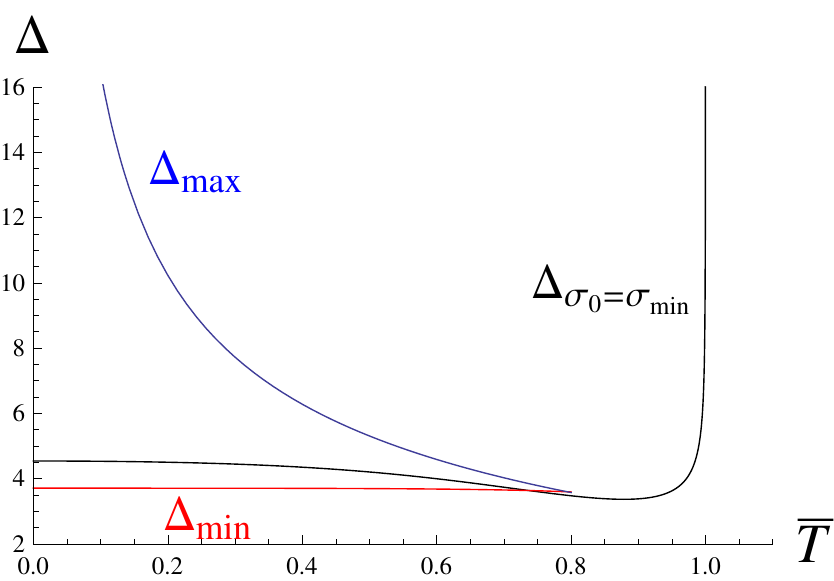}
\caption{$\Delta_{\rm max}$ (blue curve), $\Delta_{\rm min}$ (red curve), $\Delta$
 at $\sigma_{\rm min}$ (black curve) as a function of the temperature $\bar T$ for $\kappa=1$.}
\label{deltaT}
\end{figure}

By looking at the curves in Figures~\ref{DeltaPlot41} and \ref{DeltaPlot25} we can see that there are new interesting features with respect to the zero temperature case displayed in Figure \ref{DeltaCM}.
We first notice that $\sigma_0$ is bounded from below. Indeed, we found in Section \ref{sec:solution} the bound \eqref{sigmamin}
\begin{equation}
\label{sigmaminagain}
\sigma_0 \geq \sigma_{\rm min} \equiv \frac{\bar{T}^2}{(1- \bar{T}^8)^{1/4}}
\end{equation}
This is in contrast to the zero temperature case where one can take $\sigma_0 \rightarrow 0$ corresponding to the infinite spike solution.
For most values of $\bar{T}$ we further have the feature that when increasing $\sigma_0$ from $\sigma_{\rm min}$ then $\Delta$ increases until it reaches a local maximum denoted by $\Delta_{\rm max}$. Increasing $\sigma_0$ further $\Delta$ decreases until it reaches a local minimum that we denote $\Delta_{\rm min}$ (for most values of $\bar{T}$ this is also the global minimum). Increasing $\sigma_0$ further $\Delta$ increases monotonically and follows increasingly closely the zero temperature value of $\Delta$ as a function of $\sigma_0$.
In Figure \ref{deltaT} we have displayed the behavior of $\Delta$ at $\sigma_0=\sigma_{\rm min}$, $\Delta_{\rm max}$ and $\Delta_{\rm \min}$ for all values of $\bar{T}$. We see from this that there is a critical value of $\bar{T}$ given by $\bar{T}_b \simeq 0.8$ beyond which $\Delta_{\rm max}$ and $\Delta_{\rm min}$ ceases to exist. Note also that just before $\bar{T}$ reaches $\bar{T}_b$ one has that $\Delta$ at $\sigma_0=\sigma_{\rm min}$ is smaller than $\Delta_{\rm min}$, unlike for lower values of $\bar{T}$ where $\Delta_{\rm min}$ is the global minimum.

We can conclude from the above that the there is no direct analogue of the infinite spike solution with
$\sigma_0=0$ and $z(\sigma_0)=\infty$ for non-zero temperature. As one can see from \eqref{sigmaminagain} $\sigma_0$ can not reach zero for non-zero temperature. Moreover, for $\bar{T} \leq \bar{T}_b$ the local maximum $\Delta_{\rm max}$ is always finite, as one can see from Figure \ref{deltaT}. Thus, there are no finite values of $\sigma_0$ for which $\Delta $ is infinite. However, in our paper \cite{forthcoming} we examine a different way to find an analogue of the infinite spike solution at finite temperature.

The plots in Figs.~\ref{DeltaPlot41}, \ref{DeltaPlot25} and \ref{deltaT} illustrate a number of interesting features:
\begin{itemize}
\item  We observe that for any temperature, for sufficiently large $\sigma_0$,  the finite temperature curve is increasingly close to the
    corresponding extremal curve as $\sigma_0$ increases. In particular, in accordance with expectations, as the temperature is lowered a larger part of each of the two curves is close to each other. However, for any non-zero temperature $\bar{T} < \bar{T}_b$ there is always
 the new branch connected to $\sigma_{\rm min}$
going up to $\Delta_{\rm max}$.
\item As the temperature approaches its maximum value ($\bar{T} \rightarrow 1)$ the thin throat branch diminishes, and the part of the curve that
    remains coincides increasingly with that of the extremal one.
\item Because of the appearance of a maximum $\Delta_{\rm max}$ and
minimum $\Delta_{\rm min}$ for temperatures in the range $0 < \bar{T} < \bar{T}_b$, there exist values of the brane
separation $\Delta$ for which there are three possible phases. We will
compare these in more detail in Ref.~\cite{forthcoming}.
\end{itemize}

\subsection{Analytical results}

We now consider what we can say analytically about the $\Delta$ versus $\sigma_0$ behavior either in the small temperature regime or the large $\sigma_0$ regime using the analysis of Section \ref{sec:extrbranch}
for the branch connected to the extremal configuration%
\footnote{Corresponding analytical results can be obtained
for the branch connected to the neutral 3-brane, using App.~\ref{sec:neutbranch}, but these will not be discussed in this paper.} and the results listed in  App.~\ref{app:DeltaTexp}.

We begin by considering the $\Delta$ versus $\sigma_0$ behavior for small temperatures $\bar{T} \ll 1$. We first consider the minimum $\Delta_{\rm min}$ of the $\Delta( \sigma_0)$ curve (for small temperatures this is a global minimum). The small temperature expansion for $z'(\sigma)$ in \eqref{zsmallT} can be used
to compute the small temperature expansion of the brane separation \eqref{braneseparation}
in the corresponding wormhole configuration. The result is given in App.~\ref{app:DeltaTexp}, from which
we quote
\begin{equation}
\label{shortexp}
\Delta = \Delta_0 + \bar{T}^4 \Delta_1 + \CO ( \bar{T}^8 )
\end{equation}
Here the leading term  $\Delta_0$ is the extremal result \eqref{separation} (with $\kappa=1$) and
$\Delta_1$ is given in \eqref{Delta1}. Using now \eqref{shortexp} we find that the minimum is found at
\begin{equation}
\label{sigmadeltamin}
\begin{array}{c} \ds
\sigma_0 = 1+x\, \bar{T}^4 + \CO( \bar{T}^8 )
\\[2mm] \ds
     x\equiv\frac{1}{3 \sqrt{6}}+\frac{\Gamma \left(-\frac{1}{4}\right)^3 \Gamma \left(\frac{1}{8}\right)^2}{1024\,\emph{} 2^{3/4} \sqrt{3} \pi^{5/2}}
     +  \frac{\sqrt{6} \Gamma \left(-\frac{1}{4}\right) \Gamma \left(\frac{3}{4}\right) \Gamma \left(\frac{13}{8}\right)}{5 \pi  \Gamma \left(\frac{1}{8}\right)}
   \sim -0.103
   \end{array}
\end{equation}
Note that the leading result $\sigma_0=1$ reproduces the one found in Section \ref{sec:extremalwormhole}. The corresponding value of the minimum separation distance is
\begin{equation}
\Delta_{\rm min}=
\frac{2 \sqrt{2 \pi }\,\Gamma \left(\frac{5}{4}\right) }{\Gamma
   \left(\frac{3}{4}\right)}- \bar{T}^4 \,\frac{\sqrt{2 \pi }  \Gamma \left(\frac{5}{8}\right)}{3 \sqrt{3}\,\Gamma
   \left(\frac{1}{8}\right)}     +    \CO ( \bar{T}^8 )
\label{Deltamin}
\end{equation}
showing that for small temperatures the minimum is slightly lower than for $T=0$.

We turn now to the local maximum $\Delta_{\rm max}$ for small temperatures. Using the expansion \eqref{DeltaexpT12}  of $\Delta$ in powers of $\bar{T}^4$ up to and including the $\bar{T}^{12}$ term we find that the $\sigma_0$ value at which the local maximum occurs is
\begin{equation}\label{Max_sigmahat}
    \sigma_0 = a_1\, \bar{T}^{2/3}+ a_2\, \bar{T}^{10/3} +  \CO( \bar{T}^{18/3} )
\end{equation}
with $a_1$ and $a_2$ given numerically by $a_1\simeq0.693$ and $a_2\simeq-0.00243$.%
\footnote{To order $\bar{T}^{4}$ in the expansion \eqref{DeltaexpT12} we find $a_1 \simeq 0.707$ and $a_2 \simeq 0.0908$. Going to order $\bar{T}^{12}$ one instead finds $a_1\simeq0.693$ and $a_2\simeq-0.00243$.}
Note that the leading scaling $\sigma_0 \propto \bar{T}^{2/3}$ in \eqref{Max_sigmahat} for the position of the maximum
is consistent with the fact that it should be in between $\sigma_{\rm min}$, which goes like $\bar{T}^2$ to leading order, and the $\sigma_0$ value in \eqref{sigmadeltamin} corresponding to minimum $\Delta_{\rm min}$ which goes like $\bar{T}^0$ to leading order. The local maximum corresponding to \eqref{Max_sigmahat} is
\begin{equation}
\Delta_{\rm max}  \simeq  \frac{b_1}{\bar{T}^{2/3}} \, ,
\end{equation}
where $b_1 \simeq 3.28$ as computed using the expansion \eqref{DeltaexpT12} to order $\bar{T}^{12}$.
Note furthermore that the contributions to each term in the expansion
of $\sigma_0$ in \eqref{Max_sigmahat} come from each term of the
 expansion of $\Delta$ in \eqref{DeltaexpT12}. This is the reason why
we expand $\Delta$ up to $\bar{T}^{12}$ instead of considering only the first two terms (\textit{i.e.} up to the order $\bar{T}^4$) as we did in the determination of the minimum.

We can also study what happens for $\sigma_0$ large. Here we can use \eqref{zprimelargesigma}
and in an expansion for small temperature we can compute the first correction, yielding
\begin{equation}
\Delta_{\sigma_0\to\infty}=   \frac{\sqrt{\pi } \Gamma \left(\frac{5}{4}\right)}{ \Gamma \left(\frac{3}{4}\right)}\left[2\sigma_0+\frac{\kappa^2}{\sigma_0^3}\left(1-\bar{T}^4\, \frac{1}6\sqrt{3}\right) + \CO(\bar{T}^8)\right]
\label{largeDelta2}
\end{equation}
We see that for large $\sigma_0$ the leading behavior  of $\Delta$ is linear in $\sigma_0$ as in the extremal case
(see Eq.~\eqref{largeDelta}).

\section{Conclusions}
\label{sec:concl}

In this paper we have proposed a new method for D-brane probes in thermal backgrounds, studying in particular the thermal generalization of the BIon
solution. To address this construction  we have used the recently
developed blackfold approach
\cite{Emparan:2009at,Emparan:2007wm,upcoming}. While this method was originally conceived and applied in connection with the approximate analytic construction of novel black hole solutions of Einstein gravity and supergravity in five and more dimensions, the results of this paper illustrate that it has a far broader range of applicability.

In particular, the applications to new stationary blackfold solutions considered so far have mainly focused on black holes with compact horizons, and many new possible horizon topologies have been found. However, the approach is perfectly suited as well to describe the bending of black branes into other types of geometries. Consequently, it is the appropriate starting point to describe what happens to BIons when we switch on the temperature. Since the latter corresponds in the supergravity picture to an extremal D3-F1 probe brane system curved in a flat space background \cite{Lunin:2007mj} 
it is natural to base the thermal generalization on a {\it non-extremal} D3-F1 probe brane system curved in hot flat space.
The equilibrium conditions for such a configuration can then be computed from the  blackfold equations. Moreover, as we have shown in this paper, the latter equations are in fact the natural non-extremal generalization of the DBI EOMs,
providing an alternate (heuristic) derivation of the firmly established blackfold method.

One may wonder how the method employed in this paper relates
to previous works in which DBI configurations at finite temperature were considered. The commonly used method is to use the DBI
action in a thermal background, e.g. hot flat space or a Euclidean black hole background. However, this ignores the fact that,
as soon as one switches on the temperature, the probe itself should be replaced
by a thermal object. We believe the method developed in this paper is able to correctly describe this, as argued in more detail in the Sec.~\ref{sec:argument}.
One can speculate that this new perspective on finite temperature D-brane probes might resolve the discrepancies between gauge theory and gravity results found in~\cite{Hartnoll:2006hr, Grignani:2009ua}.

Another point worth emphasizing is that we have presented a thermodynamic action
(Gibbs free energy) which gives the extrinsic blackfold equations describing the thermal generalization of the BIon.
This action reduces to the DBI Hamiltonian in the zero temperature limit, and may hence be regarded as the finite temperature/closed string analogue of the corresponding DBI Hamiltonian. We also note that, as seen more generally
in the blackfold approach, the thermodynamic origin of the action implies
that the (mechanical) extrinsic blackfold EOMs are equivalent to requiring the first law of thermodynamics. The latter can be also seen as entropy maximization for given mass, so that this may be considered as a concrete manifestation of an entropic principle governing equations that originate from gravity (see \cite{Verlinde:2010hp}).

From the action we were able to obtain the explicit solution \eqref{thesolution} for the slope of the embedding function describing the brane profile of a thermal D3-brane configuration with electric flux ending in a throat. We showed that there are two branches of solutions, one connected to the extremal configuration and the other
connected to the neutral black 3-brane configuration. In most of our analysis
we focused on the former branch.
We have discussed the resulting finite temperature wormhole configuration
of $N$ D3-branes and parallel anti-D3-branes connected by a wormhole with
F-string charge. We found that the finite temperature system behaves qualitatively different than its zero-temperature counterpart.
In particular, for a given separation between the D-branes and anti-D-branes, while at zero temperature there are two phases, at finite temperature
there are either one or three phases available. 

Moreover, from our results in Sec.~\ref{sec:braneseparation} it seems that for small temperature and large enough $\sigma_0$ the non-extremal BIon solution is well-approximated by the extremal BIon solution. We take this to mean that in this range using the (abelianized $U(1)^N$) 
DBI action as a probe of hot flat space could be a good approximation to our thermal D-brane probe. This would be interesting to examine further. It also illustrates that there are certain regimes where the two methods give different results which means that using our new method can change the results both quantitatively and qualitatively for certain regimes (e.g. the ``thin throat" branch of the extremal solution becomes two branches with a maximum $\Delta$).

We left two important subjects for further considerations in the forthcoming paper \cite{forthcoming}. The first subject is about the thermodynamics of the three branches of solutions that we found in Section \ref{sec:braneseparation}. This is done by comparing the free energy for the branches in the canonical ensemble. The other subject is the apparent non-existence of a thermal generalization of the infinite spike solution as also seen in Section \ref{sec:braneseparation}. For this we consider a different type of generalization by matching the supergravity solution of $k$ non-extremal strings to our thermal D3-brane configuration with electric flux ending in a throat.

In addition there are various open problems that would be interesting to pursue, on which
we briefly comment.

First of all, it would be interesting to generalize the solution of this
paper to a thermal AdS background or AdS black hole background%
\footnote{See Refs.~\cite{Caldarelli:2008pz,Armas:2010} for applications of the blackfold approach in AdS backgrounds.}. This could have potential interesting
applications in the AdS/CFT correspondence and shed light on the above-mentioned  discrepancies \cite{Hartnoll:2006hr, Grignani:2009ua}.
More generally, since our method entails a new approach for D-brane probes in thermal backgrounds, it would be interesting to revisit other previously
studied cases in which the classical DBI action is used for the D-brane probe in the thermal
background.

We have focused in this paper on the thermal generalization for the BIon
in the case of D3-branes with electric flux. The construction is readily
generalized for D$p$-branes by starting with the EM tensor of the
non-extremal D$p$-F1 brane system. It would be useful to investigate
this more general case, which could perhaps involve some qualitatively different features depending on $p$.

The non-extremal D3-F1 brane system was treated in our construction
at the probe level, ignoring backreaction effects. It would be worthwhile to try to go beyond
the probe approximation and include such effects in a perturbative expansion.
A scheme for this,  based on matched asymptotic expansion, has been developed within the blackfold approach  and already been successfully applied to specific classes of black objects \cite{Emparan:2009at,Emparan:2007wm,Caldarelli:2008pz}. Furthermore, the blackfold approach has also given a powerful tool to study time evolution and stability \cite{Emparan:2009at,Camps:2010br}. Considering those for the
thermal generalization of the BIon would be interesting as well.

A more ambitious open direction would be to find a first principles
derivation from string theory
of the action describing thermal D-brane probes, generalizing the
DBI action derivations of \cite{Fradkin:1985qd}.  This would already be
interesting to attempt to do for the case studied in this paper, for
which we have found a thermodynamic action.
We also point out that the analysis of this paper has been at tree-level in string theory.
It might be possible to examine how the conclusions are affected by
including one (or higher)-loop effects.

Finally, we note that our method is not confined to D-branes but can be used more generally for all types of brane probes in thermal backgrounds, for example one could study M-brane probes in M-theory or NS5-brane probes in string theory.

\section*{Acknowledgments}

We thank Jan de Boer, Nadav Drukker, Vasilis Niarchos, Gordon Semenoff, Larus Thorlacius, Donovan Young and in particular Roberto Emparan
for useful discussions. TH thanks NBI for hospitality and MO and NO are grateful to Nordita for hospitality
during the workshop on ``Integrability in String and Gauge Theories; AdS/CFT Duality and its Applications''.

\appendix

\section{Analysis of branch connected to neutral configuration}
\label{sec:neutbranch}

In this appendix we discuss some relevant limits of the
solution branch \eqref{branch2} which is connected to the
neutral 3-brane.

We begin by considering the solution for small temperature.
In this case, we find that in the limit of small temperatures we have
\begin{equation}
\cosh^2 \alpha = 1 + \frac{4}{27} \cos^2 \delta + \frac{16}{243} \cos^4 \delta + \CO( \cos^6 \delta )
\end{equation}
so that after some algebra
\begin{equation}
-z' (\sigma)= \frac{\sigma_0^2}{\sqrt{\sigma^4 - \sigma_0^4}} \left( 1 + \frac{8}{27} \frac{\kappa^2}{\sigma_0^4} \bar{T}^8  \right)
\end{equation}
with the next correction being of order $\bar{T}^{16} ( 1 + \kappa^2 / \sigma_0^4 )^2$. The corresponding brane separation for the wormhole configuration
then becomes
\begin{equation}
\Delta = \frac{2\sqrt{\pi} \Gamma (5/4) }{ \Gamma(3/4)} \sigma_0 \left( 1 + \frac{8}{27} \frac{\kappa^2}{\sigma_0^4} \bar{T}^8  \right)
\end{equation}

It is also interesting to consider here what happens when the D3-brane charge
goes to zero, i.e. we have a neutral 3-brane with $N=0$. In some sense,
this can be viewed as the opposite limit of the extremal limit connected
to the first branch, which we examined above.
From \eqref{defkappa} we see that as $N \rightarrow 0$ we have that $\kappa
\rightarrow \infty$, so that from \eqref{cosdelta} we find
\begin{equation}
\label{newdefdelta}
\cos \delta (\sigma) \equiv \frac{\hat{T}^4}{\sigma^2}
\end{equation}
where we have defined
\begin{equation}
\hat T^4 = \bar T^4 \kappa = \frac{9\pi}{16\sqrt{3}} \frac{k T_{\rm F1} T^4}{ T_{\rm D3}^2}
\end{equation}
Note that in the final expression the $N$-dependence has canceled out,
after substituting the definitions of $\kappa $ in \eqref{defkappa} and
$\bar T$ in \eqref{Tbar}.
From \eqref{newdefdelta} we immediately read off the lower bound on $\sigma$ (and hence $\sigma_0$) for a given temperature
\begin{equation}
\sigma_{\rm min} = \hat{T}^2
\end{equation}
Substituting now \eqref{newdefdelta} in the the relevant solution \eqref{branch2} for $\cosh \alpha$ and using this in \eqref{zprimesolution} we obtain
\begin{equation}
-z' (\sigma)= \frac{\sigma_0^2}{\sqrt{\sigma^4 - \sigma_0^4}} \left( 1 + \frac{8}{27} \frac{\hat{T}^8}{\sigma_0^4}   \right)
\end{equation}
From this we then find the corresponding brane separation as
\begin{equation}
\Delta = \frac{2\sqrt{\pi} \Gamma (5/4) }{ \Gamma(3/4)} \sigma_0 \left( 1 + \frac{8}{27} \frac{\bar{T}^8}{\sigma_0^4}   \right)
\end{equation}

\section{$\Delta$ for small temperatures}\label{app:DeltaTexp}

Here we show explicitly the small temperature expansion of
$\Delta$, \textit{i.e.} the separation between the two systems of D3-branes.
Using
\begin{equation}
	\Delta = 2 \int_{\sigma_0}^\infty d\sigma \left( \frac{F(\sigma)^2}{F(\sigma_0)^2} - 1 \right)^{- \frac{1}{2}} Ê\, ,
\end{equation}
along with \eqref{Ffunction} and \eqref{mainbranch} we can expand $\Delta$ for small $\bar{T}$ up to the order
$\bar{T}^{12}$
\begin{equation}
\Delta=\Delta_0 + \bar{T}^4\Delta_1+\bar{T}^8\Delta_2+\bar{T}^{12}\Delta_3+\mathcal{O}(\bar{T}^{16}) \,
\label{DeltaexpT12} \, ,
\end{equation}
where $\Delta_0$ is given in equation \eqref{separation} and $\Delta_1$, $\Delta_2$, $\Delta_3$ are computed to be
\begin{equation}
\label{Delta1}
	\frac{\Delta_1}{\sqrt{\kappa}}=\frac{2 \sqrt{\frac{\pi }{3}} Ê\Gamma \left(\frac{5}{4}\right)}{3 y^7
 Ê \Gamma \left(-\frac{1}{4}\right)} \left(y^4+1\right) \left[- y^2 \sqrt{y^4+1}\, {}_2F_1\left(-\frac{1}{2},\frac{5}{4},-\frac{1}{4};-\frac{1}{ y^4}\right)-1+ y^4\right]
\end{equation}
\begin{equation}
\begin{split}
\frac{\Delta_2}{\sqrt{\kappa}}&=\frac{\sqrt{\pi } \Gamma \left(\frac{1}{4}\right) \sqrt{y^4+1}}{648
 Ê y^{13}  \Gamma \left(-\frac{1}{4}\right)}
 Ê \left[5 y^2 \left(y^4+1\right)^{3/2}\left(2\left(2
 Ê y^8+5 y^4 \right)\, {}_2F_1\left(-\frac{1}{2},\frac{1}{4},-\frac{5}{4};-\frac{1
 Ê }{y^4}\right)\right.\right.\cr&\left.\left.-\left(y^4+1\right) \left(4
 Ê y^4+7 \right)\, {}_2F_1\left(-\frac{1}{2},\frac{5}{4},-\frac{5}{4};-\frac{1
 Ê }{y^4}\right)\right)-3 1 \left(y^{12}-6 y^8
 Ê +11 y^4 +10 \right)\right]
\end{split}
\end{equation}
\begin{equation}
\begin{split}
		\frac{\Delta_3}{\sqrt{\kappa}}&=\frac{\sqrt{\frac{\pi }{3}}\Gamma \left(\frac{1}{4}\right)
 Ê \left(y^4+1\right)}{22680 y^{19} 
 Ê  \Gamma \left(-\frac{5}{4}\right)} \left[ 7  \left(-42
 Ê y^{16}-y^{12} +65 y^8+145
 Ê y^4 +105 \right) 
\right.\cr&\left.
 Ê\,
-6 y^6
 Ê \left(y^4+1\right)^{3/2} \left(171 y^{12}+326
 Ê y^8 +294 y^4 +385 \right) {}_2F_1\left(-\frac{1}{2},\frac{1}{4},-\frac{9}{4};-\frac{1}{y^4}\right)
\right.\cr&\left. Ê\,
+3 y^2
 Ê \left(y^4+1\right)^{5/2} \left(342 y^{12}+386
 Ê y^8 +504 y^4 +385 \right)
 {}_2F_1\left(-\frac{1}{2},\frac{5}{4},-\frac{9}{4};-\frac{1}{y^4}\right)\right]
\end{split}
\end{equation}
where we defined $y\equiv \sigma_0 / \sqrt{\kappa}$ and ${}_2F_1(a,b,c;z)$ is the hypergeometric function.


\addcontentsline{toc}{section}{References}




\providecommand{\href}[2]{#2}\begingroup\raggedright\endgroup

\end{document}